\documentclass [12pt,twoside]{article}           % version LATEX2E
\usepackage[left,modulo]{lineno}
\usepackage{setspace}
\countdef\arxiv=201
\ifnum\arxiv=0 \input cpreb.tex \fi

\ifnum\arxiv=1

\usepackage{epsfig,times,lscape}
\usepackage[usenames]{color}

    \ifnum\count102=1

    \fi

    \ifnum\count102=2
\fi

\newcommand{\nc}{\newcommand}

     \ifnum\count101=1
\nc{\qI}[1]{\section{{#1}}}
\nc{\qA}[1]{\subsection{{#1}}}
\nc{\qun}[1]{\subsubsection{{#1}}}
\nc{\qa}[1]{\paragraph{{#1}}}

\def\qpar{\vskip 2mm plus 0.2mm minus 0.2mm}
\def\qL{\hfill \break}
     \fi 

      \ifnum\count101=2
 \nc{\qI}[1]{\parindent=0mm \vskip 8mm 
{\centerline{\LARGE \color{red}#1}}\vskip 3mm}
\nc{\qA}[1]{\vskip 2.5mm \noindent 
{{\bf\large\color{blue}  #1}} \vskip 1mm \parindent=0mm}
 \nc{\qun}[1]{\vskip 1mm \noindent {\sl #1 }\quad }

\def\qL{\hfill \break}
\def\qpar{\vskip 2mm plus 0.2mm minus 0.2mm}

      \fi
\def\qth{\vrule height 12pt depth 0pt width 0pt}
\def\qtb{\vrule height 0pt depth 5pt width 0pt}

\def\qtH{\vrule height 20pt depth 0pt width 0pt}
\def\qtB{\vrule height 0pt depth 10pt width 0pt}
\nc{\qfoot}[1]{\footnote{{#1}}}

      \ifnum\count101=1
\def\qbu{\hfill \par \hskip 6mm $ \bullet $ \hskip 2mm}
\def\qee#1{\hfill \par \hskip 6mm (#1) \hskip 2 mm}
      \fi
      \ifnum\count101=2
\def\qbu{\hfill \par \hskip 4mm $ \bullet $ \hskip 2mm}
\def\qee#1{\hfill \par \hskip 4mm (#1) \hskip 2 mm}
      \fi

\def\qparr{ \vskip 1.0mm plus 0.2mm minus 0.2mm \hangindent=10mm
\hangafter=1}

     \ifnum\count101=1 
 
     \fi
     \ifnum\count101=2

  \def\qcitb#1{\noindent \hbox to 102mm{\hfill \small #1} \vskip 1mm}
      \fi

 \def\qpages#1{\count102=0{\loop\advance\count102 by 1
 \null \vfill\eject \ifnum\count102<#1 \repeat}}

\def\qth{\vrule height 12pt depth 0pt width 0pt}
\def\qtb{\vrule height 0pt depth 5pt width 0pt}

\def\qv{\vskip 0.1mm plus 0.05mm minus 0.05mm}

\def\qhw{\hskip 1.5mm}
\def\qleg#1#2#3{\noindent {\bf \small #1\qhw}{\small #2\qhw}{\it \small #3}\qv }
\fi
\begin{document}
\thispagestyle{empty}

% --------------------------------------------------------------------

      % Hauts de pages et numerotation

          % Remarque: sans le \protect --> message d'erreur (ordre fragile)
\markboth{{\sl \hfill  \hfill \protect\phantom{3}}}
        {{\protect\phantom{3}\sl \hfill  \hfill}}

% -------------------------------------------------------------------
\color{yellow} 
\hrule height 20mm depth 10mm width 170mm 
\color{black}
\vskip -2.2cm 

 \centerline{\bf \Large Effect of marital status on death rates.}
\vskip 2mm
 \centerline{\bf \Large Part 1: High accuracy exploration of the 
Farr-Bertillon effect. }
\vskip 18mm
\centerline{\large 
Peter Richmond$ ^1 $ and Bertrand M. Roehner$ ^2 $
}

\vskip 5mm
\large

{\bf Abstract}\quad
The Farr-Bertillon law says that for all age-groups the death
rate of married people is lower than the death rate of people
who are not married (i.e. single, widowed or divorced).
Although this law has been known for over 150 years, it
has never been established with great accuracy. This even let some
authors argue that it was a statistical artifact.
It is true that the data must be selected and
analyzed with great care, especially
for age groups of small size such as widowers under 25. \qL
The observations reported in this paper were selected and
designed in the same way as experiments in physics, that is to
say with the objective of minimizing the error bars 
for all age-groups. 
It will be seen that data
appropriate for mid-age groups may be unsuitable for young
age groups and vice versa. 
\qL
The investigation led to the following results.
(1) The FB effect is basically the same for men and women,
except that on average it is about 20\% stronger for men.
(2) There is a marked difference between single or divorced
persons on the one hand,
for whom the effect is largest around the age of 45,
and widowed persons on the other hand,
for whom the effect is largest around the age of 25.
(3) When different causes of death are distinguished, 
the effect is largest for suicide and smallest for cancer.
(4) For young widowers the death rates are up to 10 times
higher than for married persons of same age.
This extreme form of the FB 
effect will be referred to as the ``young widower
effect''.\qL
A possible connection between the FB effect and Martin Raff's ``Stay
alive'' effect for cells in an organism is discussed in the last section.

\vskip 5mm
\centerline{\it Version of 19 August 2015. Comments are welcome.}

\vskip 1mm
{\normalsize Key-words: death rate, marital status, widowhood, young
widowers, apoptosis.}

\vskip 2mm

{\normalsize 
1: School of Physics, Trinity College Dublin, Ireland.
Email: peter\_richmond@ymail.com \qL
2: Institute for Theoretical and High Energy Physics (LPTHE),
University Pierre and Marie Curie, Paris, France. 
Email: roehner@lpthe.jussieu.fr
}

\vfill\eject

\large

\qI{Introduction}

Let us first define several terms which will be used
throughout this article.
\qbu The marital status of a person refers to one of the
following situations: single, married, widowed, divorced.
Needless to say, ``single'' means that the person has {\it never}
been married for otherwise he (or she) would be
widowed or divorced. These groups will be designated
by the letters $ s,m,w,d $ respectively. The case of people
who are married but separated or not married but cohabiting
will also be considered later on albeit fairly shortly.
\qbu For each of these groups of persons one can define
a death rate in the standard way, that is to say
by dividing the number of persons who die annually
by the size of the group. In addition to the marital
status distinction, one can order people by age group.
For instance, $ d_m(15:24) $ will be the death rate
of married persons who are between 15 and 24 year old.
\qbu Finally, we introduce the notion of {\it death
rate ratio} which is the death rate of a given group
divided by the death rate of married persons of same age.
For instance, the death rate ratio of widowed persons
in the age group $ 15:24 $ will be:
$$ \hbox{Death rate ratio of widowed persons:}\quad
 r_w(15:24) = d_w(15:24)/d_m(15:24) $$

The expression {\it death rate ratio distribution of 
widowed persons} 
will refer to the curve of $ r_w $ as a function of age.
Sometimes, death rate ratio distributions
will also be named Farr-Bertillon distributions.

\qA{The Farr-Bertillon law}

In the social sciences there are very few laws which are valid
at any time and in any country. The Farr-Bertillon law%
\qfoot{So far, in the literature 
the FB law was variously referred to as
the ``marriage effect'' , the ``widower effect'' or the
``bereavement effect''. Adding to the confusion, 
some of these expressions were meant to describe special
facets; for instance the term ``bereavement effect'' focuses
on short-term rather than permanent effects.
Here, as is standard in physics, this law
will be designated by the name of its discoverers. We hope
that following this usage will clarify its significance.}
which states that for all age-groups 
married persons have a lower death rate than unmarried persons
is one of them. 
More precisely, in all cases for which reliable data
are available this law holds with error bars which are not
broader than $ \pm 10\% $. 
\qpar
At first sight, our assertion that there
are few laws of this kind may seem surprising. For instance,
is it not true that the frequency distribution of high incomes
follows a Pareto law?  Compared with the Farr-Bertillon law
there are two major differences, however.
\qbu The Pareto law contains a free parameter, namely the
exponent of the power law. The Farr-Bertillon effect
contains no free parameter.
\qbu The Pareto law describes a frequency distribution whereas
the Farr-Bertillon law is a relationship between two
``physical'' variables. In short, the Pareto law is of the same kind
as the Maxwell-Boltzmann law which gives the velocity
distribution of the molecules of a gas 
whereas the
Farr-Bertillon law 
is similar (for instance) to Einstein's law which gives the
relationship between specific heat and temperature.  
Needless to say, a relationship between physical variables
tells us more about the system than a probability distribution%
\qfoot{The MB distribution for the speed
of molecules is a consequence of the fact that
each velocity component follows a centered Gaussian distribution.
Because of the central limit theorem, Gaussian distributions are
very common in the natural sciences which means that the
exponential shape of the MB distribution only tells us that it
belongs to this broad class rather than to
the power law class. Actually, all significant 
physical information (about molecules masses
and temperature) is contained in the width of the MB distribution.}%
.
\qparr

The Farr-Bertillon law consists in the fact that married people
have lower death rates than non-married people, either never married,
widowed or divorced people. It is named after
William Farr (1807--1883) and Louis-Adolphe Bertillon (1821--1883).
In 1859 Farr observed the effect on French data. Both
Farr and Bertillon were among the main founders of medical demography.
Bertillon's strong focus on comparative international 
investigations led him to recognize the existence of
this effect in a broad range of countries (Bertillon 1872).
As a matter of fact, in the one and a half century since its
discovery,
the Farr-Bertillon effect has been observed
in {\it all} countries for which reliable data are available.

\qA{Measurement issues}

Why did we emphasize that the data must be reliable?\qL
Even today in industrialized countries
it remains a real challenge to produce the data which are
needed to observe this law. It can even be said that present-day
data are probably less accurate than those of 50 or 100 years
ago for, as will be seen later, present-day statistics rely
more and more on surveys based on population samples.
\qpar
 
Why is it a difficult task to measure the death rate of widowed
persons and particularly of young widowers?
As all death rates, the death rate $ d(G) $
of widowers in age-group $ G $
is defined as a ratio:
$$ d(G)=
{ \hbox{Number of widowers who died during the year: }D(G) \over
  \hbox{Population of widowers at beginning of year 
in given age group: }P(G)
} $$

The numerator is fairly easy to measure because in all countries
age and 
marital status are two characteristics recorded on death
certificates. In contrast, it is difficult to
get reliable estimates for $ P(G) $. 
There are (at least) 4 difficulties in measuring $ d(G) $ for
young widowers.
\qee{1} In the age groups under 30 there are few widowers.
For instance in 1980 in the United States the 15-19 age group
had only 6,448 widows and 2,081 widowers%
\qfoot{There are less widowers than widows because at the same age
there are less married men than married women.}%
.
In the 15-24 age group there were 31,100 widows and 8,050 widowers. 
In any census the task of
identifying and counting accurately such small populations 
is not easy. Naturally, if instead of real censuses one relies
on sample surveys, the task becomes even more difficult
or altogether impossible if the samples are too small.
\qee{2} Even though in a general way it is easier to
measure death numbers than to count living persons, 
the fact that there are very few deaths of young widowers 
creates huge statistical fluctuations.
Thus, in the United States in 1980 for the age-group 15-24
there were only 38 deaths of widows and 29 deaths of widowers.
In the following age group 25-34 the numbers were about 
ten times larger, 323 and 219 respectively, but these are
still small numbers. As a matter of fact, the deaths of widowed
persons become ``substantial'' only over 65 years of age. Thus,
in the age group 65-74 there were 35,630 deaths of widowers. 
\qee{3} $ P(G) $ must be measured through a census but the problem
with censuses is that they are based on the answers provided by the
respondents. Even in countries such as the United States where
censuses have been organized with much care, the enumerator relies
entirely on the answers provided by the head of the household%
\qfoot{In fact, after 1990 the census forms were mailed
to the persons; visits by enumerators were limited
to a few households.}%
.
Yet, it is well known that the answers provided by the respondents
are not always accurate. For instance, even such basic variables
as age or the number of years spent in widowhood
are not well remembered especially by elderly people.
It appears that often such variables
are rounded up to the nearest multiples of 10 or 5.
That is why most census forms ask both the age and
the year of birth. 
Moreover, the answers may be affected by other forms of bias.
Thus, people may prefer to say that they
are widowed rather than separated or divorced.
\qee{4} In the interval between census years, most national statistical 
institutes carry out surveys based on population samples.
The quality of such surveys greatly depends upon how well the
samples are selected. In the United States, the annual ``Current
Population Reports'' (CPR) are based on samples of 
some $ 60,000 $ persons,
that is to say one per 20,000. Thus, even if the sample was
selected properly, there will be substantial sampling errors.
For a population of the order of
100,000 the sampling error of the CPR of 1960 was 37,000 that
is to say nearly 40\%; for a population of one million the
standard error was still 12\%
(Accuracy 1960, p. 6, Table D). As a result, good measurements
of $ d(G) $ for young widowers can be obtained only in census years.
A more detailed analysis is given below in Table 3b.

\qA{Farr-Bertillon effect in the 19th century}

In this subsection and in the next we show some of the results
due to Louis-Alphonse Bertillon. As already observed, contrary
to William Farr, Bertillon was a comparativist. After having
identified this effect in France, his main concern was to
see if it was also present in other countries for which 
data were available.

%%-----------------------------------------------
\begin{figure}[htb]
\centerline{\psfig{width=15cm,figure=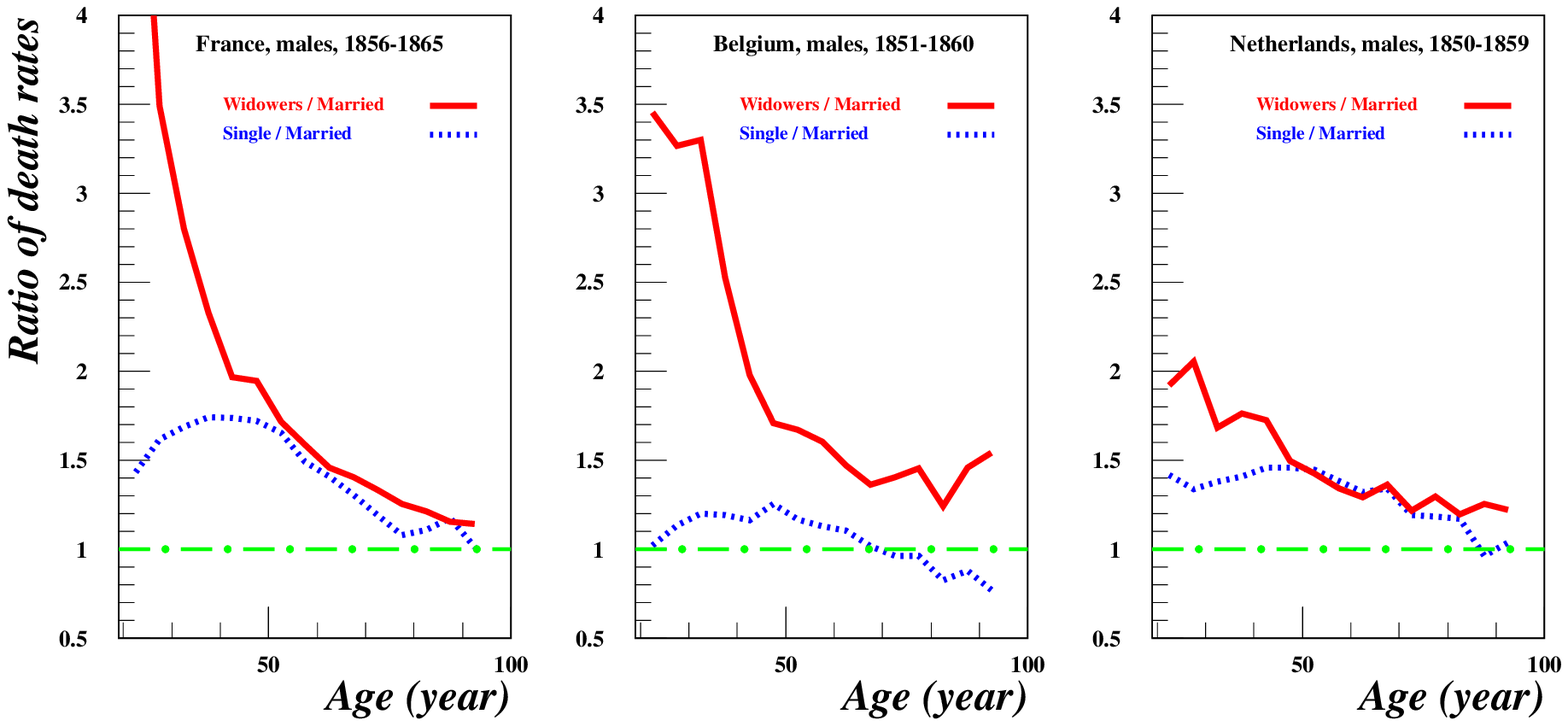}}
\qleg{Fig. 1a,b,c: Death rate ratio according to marital status.}
{The data points correspond to 16 age groups ranging from
20-25 to 95-100.
It can be seen that
the ratio widowers/married differs from the ratio single/married
both in shape and in magnitude. Incidentally, Bertillon was suspecting
a possible statistical bias for the death rates of single persons
in Belgium because they are two thirds the size of
those in France and the
Netherlands. In this graph as in the rest of the paper ``single''
means ``never married''. }
{Source: Bertillon (1872); the graph for
France appears also in Bertillon (1879, p. 781)}
\end{figure}
%-------------------------------------------------

\qA{Bonds between parents and children}
If the bond between husband and wife plays a role in the FB effect
it seems plausible to expect
the ties between parents and their
children to have a similar effect.
This conjecture is confirmed by the data in Table 1.

%%--------------------------------------------------
\begin{table}[htb]
 \centerline{\bf Table 1: Effect of
husband-wife ties and parents-children ties on suicide rates.}

\vskip 3mm
\hrule
\vskip 0.7mm
\hrule
\vskip 2mm

\color{black} 

$$ \matrix{
\qtb
\hbox{Situation} \hfill  & \hbox{M} & \hbox{F} &\hbox{ }&\hbox{M} & \hbox{F}\cr
\noalign{\hrule}
\qth 
\hbox{Married with children} \hfill& 20 & 4.5& & 1 & 1\cr
\hbox{Married without children} \hfill& 47 & 16 & & 2.4&3.6 \cr
\hbox{Widowed with children} \hfill& 52 & 10 & & 2.6& 2.2\cr
\qtb
\hbox{Widowed without children}\hfill & 100 & 23 & & 5.0 & 5.1\cr
\noalign{\hrule}
} $$
\vskip 0.5mm
Notes: The table gives average suicide rates (per 100,000 people)
in France over the 8-year time interval between 1861 and 1868. 
``M'' means male and  ``F''  female.
The two columns on the right-hand
size repeat the same data with a normalization 
based on the  situation ``married with children''.
If one accepts the explanation introduced by Emile 
Durkheim (1897) that it is the severance (or lack)
of bonds and especially of family bonds which is the main 
factor in the phenomenon of suicide, then these data allow
us to compare the respective strengths of the bonds 
between husband and wife on the one hand 
and between parents and children on the other hand.
The fact that the suicide rate is almost the same
for married persons without children as for widowed 
persons with children
suggests that the parents-children and husband-wife
bonds are of same strength.
\qL
{\it Source: Bertillon (1879, p. 474)}
\vskip 3mm
\hrule
\vskip 0.7mm
\hrule
\end{table}
%%-----------------------------------------------

\qI{Toward accurate observations of the Farr-Bertillon law}

\qA{Mid-age groups versus young age groups}

In the previous section we emphasized the fact that the
Farr-Bertillon effect holds with a level of precision akin
to what one is used to in the natural sciences. 
However, in order to reduce the error bars as much as possible
an appropriate methodology must be be used.
In this respect 
age-groups over 35 and age-groups under 35 will require
different techniques.
\qbu To estimate the sizes of the age groups over 35 one does
not necessarily need to use censuses. Estimates from surveys
based on population samples may be sufficient at least
if the samples are ``not too small''. This will allow
observations over time intervals containing a substantial
number, $ n $, of inter-census years. For the averages computed
over such time intervals, the error bars will be reduced by
the standard $ 1/\sqrt{n} $ factor. 
\qbu On the contrary, in the investigation of the young widower
effect one needs to focus on age groups under 35 and,
as already mentioned, this requires to rely on population
data from decennial censuses. To some extent, the procedure
based on census data is also needed for elderly age groups
over 75 because of their small size.
\qpar

In short, there will be two phases in our investigation.
In a first phase we will focus on the central part of the
age interval (30-60) and use as many years as possible to
get the smallest error bars. \qL
In the second phase, we will use accurate population
data available for only a few years.
This will give the death ratio for young age-groups.
Though this procedure will of course also provide results  
for central age-groups, they will be less precise than
those computed in the first phase.

\qA{Methodological options}

In the present paper we perform repeated observations. At first
sight one may think that they should be aggregated together.
However, as these observations are not performed under identical
conditions (see table 1) lumping them together would lead
to unpredictable and uncontrollable results. 
As explained below, this is
a widespread difficulty in the social sciences.
\qpar
As a matter of fact,
nothing is more depressing than to read sociological review
papers. Why? \qL
Most often, the authors of such papers
report conflicting  results 
obtained by different researchers but without describing the
conditions under which the observations were made and
how the data were analyzed. Inevitably, this makes readers fairly
uncomfortable. One gets the feeling of being confronted 
with a soft, multiform, shapeless and labile
world about which no
clear, univocal statement can ever be made. Needless to say,
if true, such a view would condemn any scientific investigation 
from the outset because reproducibility is a crucial requirement 
in any science.
\qpar
Let us illustrate this point through an example.
A paper by Koposawa et al. (1995) found that no
additional risk of suicide is significantly
associated with the marital status
of widowed or never-married persons. Such a
conclusion is at variance with the results
reported consistently by numerous former and subsequent
studies including the 
present one. If presented without specific explanations
about its methodology, this study would
give the impression that even the most unlikely claim can
be made and sustained. A closer look reveals that, in
contrast with most other investigations, this one
does not rely on aggregated data but on
a multivariate analysis of individual data. The sample contains
only 216 suicide cases. As such a small sample implies broad confidence
intervals it is 
hardly surprising that the study could not find any significant
connection between marital status and suicide rates. This
does not mean that the connection does not exist but rather that
the data used in this study were dominated by background noise.
\qpar

%%-----------------------------------------------

\begin{table}[htb]

\centerline{\bf Table 2: Summary of the observations of the 
Farr-Bertillon effect}

\vskip 3mm
\hrule
\vskip 0.7mm
\hrule
\vskip 1.2mm

\color{black} 
\small

$$ \matrix{
\hbox{Fig.} \hfill& \hbox{Country}\hfill & \hbox{Period}\hfill& 
\hbox{Population}\hfill& 
\hbox{Error}\hfill& \hbox{Shape}\hfill& \hbox{Quality} \cr
\qtb
\hbox{} \hfill& \hbox{}\hfill & \hbox{}\hfill& 
\hbox{estimates}\hfill& 
\hbox{bars}\hfill& \hbox{of w/m}\hfill& \hbox{stars}\hfill \cr
\noalign{\hrule}
\qtH
\hbox{1a} \hfill& \hbox{France}\hfill & 1856-1865\hfill& 
\hbox{?}\hfill& 
\hbox{no}\hfill& > &  \hbox{\Large *} \cr
\hbox{1b} \hfill& \hbox{Belgium}\hfill & 1851-1860\hfill& 
\hbox{?}\hfill& 
\hbox{no}\hfill& >& \hbox{\Large *} \cr
\hbox{1c} \hfill& \hbox{Netherlands}\hfill & 1850-1859\hfill& 
\hbox{?}\hfill& 
\hbox{no}\hfill& < & \hbox{\Large *} \cr
\hbox{2a,b,c} \hfill& \hbox{USA}\hfill & 1996-2010\hfill & 
\hbox{?}\hfill& 
\hbox{yes}\hfill& \hbox{?} & \hbox{\Large **} \cr
\hbox{3a,b,c} \hfill& \hbox{USA}\hfill & 1940,1950,1960 \hfill& 
\hbox{census}\hfill& 
\hbox{yes}\hfill& < & \hbox{\Large ***} \cr
\hbox{4} \hfill& \hbox{USA}\hfill & 1980,1990,2000\hfill& 
\hbox{census}\hfill& 
\hbox{yes}\hfill& < & \hbox{\Large ***} \cr
\hbox{5} \hfill& \hbox{USA}\hfill & 2005-2010\hfill& 
\hbox{ACS}\hfill& 
\hbox{yes}\hfill& < & \hbox{\Large ***} \cr
\hbox{6} \hfill& \hbox{USA}\hfill & 1980,1990\hfill& 
\hbox{census}\hfill& 
\hbox{yes}\hfill& >\ < & \hbox{\Large **} \cr
\hbox{7a} \hfill& \hbox{France}\hfill & 1968-1993\hfill& 
\hbox{?}\hfill& 
\hbox{yes}\hfill& & \hbox{\Large **} \cr
\qtb
\hbox{7b} \hfill& \hbox{France}\hfill & 1981-1993 \hfill& 
\hbox{?}\hfill& 
\hbox{no}\hfill& > & \hbox{\Large *} \cr
\noalign{\hrule}
}
$$
\vskip 2mm
Notes:
The column
``Population estimates'' indicates how the populations by age
and marital status have been estimated.  
ACS means ``American Community Survey''. 
An interrogation mark in this column
means that the technical notes of the publication failed
to explain how the estimates were computed.
The column ``Error bar''
indicates if it was possible to estimate the standard deviation of
the death rate ratios. 
The column ``Shape of w/m'' indicates
whether the curve for the death rate ratio of widowers is steadily
decreasing ($ > $) or shows a maximum for the second youngest
age group ($ < $). It is the last case which prevails in the
observations of highest quality. The column ``Quality stars''
gives an estimated quality index for each observation: three stars
means highest quality.\qL
It can be noted that similar death statistics by marital
status and age are also available for England (see Mortality
Statistics, review of the Registrar General, National Center for
Health Statistics 1970, Registrar General 1971) and Germany
(see Statistisches Jahrbuch 1978).
{\it }
\vskip 2mm
\hrule
\vskip 0.7mm
\hrule
\end{table}
%%-----------------------------------------------

Only observations of same nature and quality can possibly be lumped
together. Thus, in the observations listed in Table 2 it would be
possible to lump together the observations 3 and 4. However,
from 1940 to 2000
they would span a time interval of 60 years during which 
important population changes took place in the United States.
By keeping these
observations separate one can control whether or not 
there was a possible shift.  
\qpar

Before carrying out the program outlined in Table 2, some
preliminary tests are required. In the previous discussion
we said that population estimates based on surveys may
be acceptable provided that the samples are ``not too small''.
Obviously, one needs to clarify what is meant by this
expression. This will be done in the next section.

\qI{Sampling errors for population estimates}

The expression ``sampling errors'' corresponds to
measurement errors due to purely random fluctuations.
However, we will see that for some sample estimates 
there are also non-sampling errors which refer to
more or less systematic biases. As an example, one can
mention the response rate. Nowadays, once the sample has
been selected the forms are mailed to the respondents.
Not all of them will reply, however. In the United States,
response rates usually range between 80\% and 95\%.  
The persons who do not respond most likely are
``unstable'' households who move frequently and for that
reason may not have received the form 
and also elderly persons who are in hospital or nursing homes. 
\qpar

Table 3a describes US statistical sources. Their accuracy
will then be ascertained through a number of tests
performed in Table 3b.

%%-----------------------------------------------
\begin{table}[htb]

\centerline{\bf Table 3a: Statistical sources for US population
by marital status and age}

\vskip 3mm
\hrule
\vskip 0.7mm
\hrule
\vskip 1.2mm

\color{black} 
\small

$$ \matrix{
\hbox{Year} & \hbox{Source}\hfill &\hbox{Size of} \hfill\cr
\qtb
\hbox{} & \hbox{}\hfill &\hbox{sample} \hfill \cr
\noalign{\hrule}
\qth
\hbox{\bf \color{blue} Census years} \hfill & \hbox{}\hfill &\hbox{} \cr
1900-1940  \hfill& \hbox{Historical Statistics of the US (p. 20-21)}\hfill
& \hbox{Whole population} \hfill \cr
1950,\ 1960  \hfill& \hbox{Historical Statistics of the US (p. 20-21)}\hfill
& \hbox{25\% sample} \hfill \cr
1970  \hfill & \hbox{Historical Statistics of the US (p. 20-21)}\hfill
& \hbox{5\% sample} \hfill \cr
1980 \hfill& \hbox{Census volume PC80-1-D1-A}\hfill &\hbox{Whole
  population}\hfill \cr
1990 \hfill& \hbox{Census volume CP-1-1}\hfill &\hbox{Whole population}\cr
2000 \hfill& \hbox{Census table PCT007 on ``FactFinder''}\hfill &
\hbox{20\% sample}\hfill\cr
2010 \hfill& \hbox{{\it Not recorded}, replaced by ACS (see below)}\hfill
&\hbox{}\cr
\hbox{\bf \color{blue} Inter-census years} \hfill & \hbox{}\hfill &\hbox{} \cr
1901-1959 \hfill&  \hbox{\it No data are available}\hfill & \hbox{} \cr
1961-2004 \hfill&  \hbox{Current Population Reports (CPR)}\hfill
& \hbox{33,000-57,000} \hfill \cr
\qtB
2005- \hfill&  \hbox{American Community Survey (ACS) on ``FactFinder''}\hfill
& \hbox{2.5 million} \hfill \cr
\noalign{\hrule}
}
$$
\vskip 2mm
Notes: The population samples used in the CPR are much too
small to reflect widower populations under age 40 (see Table 3b).
This means that before 2005 there are in fact no appropriate data
for inter-census years.
Incidentally, it can be observed that the data for marital
status by age which are published in the annual volumes of
the ``Statistical Abstract of the United States'' are identical
to those published in the CPR (P20 Series). The only 
difference is that the age-group 17-18 is omitted.
This omission is probably motivated by the fact that 
for this age-group the estimates would be fairly poor.
However, the comparisons performed
in Table 3b show
substantial discrepancies even for older 
age groups up to $ 35-44 $. 
``FactFinder'' which is mentioned in the table refers to
a search engine for statistical tables which is available on the
website of the US Census Bureau.
{\it }
\vskip 2mm
\hrule
\vskip 0.7mm
\hrule
\end{table}
%%-----------------------------------------------

%%-----------------------------------------------
\begin{table}[htb]

\centerline{\bf Table 3b: Percentage errors in various estimates
of widower population.}

\vskip 3mm
\hrule
\vskip 0.7mm
\hrule
\vskip 1.2mm

\color{black} 
\small

$$ \matrix{
\hbox{Year} & \hbox{Source}\hfill &\hbox{Sample} & \hbox{} &
\hbox{} & \hbox{} & \hbox{} & \hbox{} & \hbox{} \cr
\qtb
& \hbox{}\hfill &\hbox{size} &  & &  &  &  &  \cr
\noalign{\hrule}
\qth
\color{blue}1980 & \hbox{}\hfill & &  & &  &  &  &  \cr
\hbox{Age}\hfill& & & 15-17 & 18-19 & 20-24  & 25-29 &30-34  & 35-39 \cr
& \hbox{Census}\hfill & 100\% & 992 &1,089 &5,970  &11,759  &16,531  & 
22,337 \cr
& \hbox{CPR}\hfill & 0.02\%& 0 & 0&2,000  &8,000  &11,000  & 19,000 \cr
& {\color{red}\hbox{\color{red}Census-CPR}\over
    \hbox{\color{red}Census} } \hfill & 
&\color{red} 100\% & \color{red}100\%& \color{red}66\% &
\color{red}32\% & \color{red}33\% & \color{red}15\% \cr
\color{blue}2000 & \hbox{}\hfill & &  & &  &  &  &  \cr
\hbox{Age}\hfill& & &  & 15-19 & 20-24 & 25-29 &30-34  & 35-44 \cr
& \hbox{Census}\hfill & 100\%& 
\hbox{} &  \hbox{n.a.}&\hbox{n.a.}&\hbox{n.a.}&\hbox{n.a.}&\hbox{n.a.} \cr
& \hbox{Census}\hfill & 20\%&  & 13,814&19,376  &19,604  &26,939  & 
106,135 \cr
& \hbox{CPR}\hfill & 0.02\%& &3,000 & 0& 9,000 & 15,000 &   96,000 \cr
& {\color{red}\hbox{\color{red}Census-CPR}\over
    \hbox{\color{red}Census} } \hfill & &
& \color{red}78\%  & \color{red}100\%& \color{red}54\% &
\color{red}44\% & \color{red}10\% \cr
\color{blue}2010 & \hbox{}\hfill & &  & &  &  &  &  \cr
\hbox{Age}\hfill& & & 15-17 &18-19 & 20-24 & 25-29 & 30-34 & 35-39 \cr
& \hbox{Census}\hfill & 100\%& 
\hbox{n.a.} &  \hbox{n.a.}&\hbox{n.a.}&\hbox{n.a.}&\hbox{n.a.}&\hbox{n.a.} \cr
& \hbox{Census}\hfill & 20\%& 
\hbox{n.a.} &  \hbox{n.a.}&\hbox{n.a.}&\hbox{n.a.}&\hbox{n.a.}&\hbox{n.a.} \cr
& \hbox{ACS}\hfill & 1\% & 825 &1,372 &4,572  & 9,199 & 15,876 & 28,757 \cr
& \hbox{ASES}\hfill & 0.05\%& 5,000 &3,000 & 3,000  & 21,000 &28,000
& 29,000 \cr
\qtB
& {\color{red}\hbox{\color{red} ACS-ASES}\over \hbox{\color{red} ACS}
  } \hfill & 
& \color{red} -506\% & \color{red} -119\% &\color{red} 34\% &
\color{red} -128\% & \color{red} -76\% & \color{red} -1\% \cr
\noalign{\hrule}
}
$$
\vskip 2mm
Notes: CPR means ``Current Population Reports'';
ACS means ``American Community Survey''; ASES means ``Annual Social
and Economic Survey''; n.a. means ``not available''.
The ratios (Census-CPR)/Census represent the errors in CPR estimates.
The ratios (ACS-ASES)/ACS can also be seen as roughly representing
the errors in ASES estimates.
The sample size is given as a percentage of the total US population.
\qL
In the 2000 census the question of marital
status was not asked on the short form sent to all people
but only on the long form filled by about 20\% of the population.
In the census of 2010 the marital status question was not asked at all. 
It was  replaced by the ACS, yet with lower accuracy due to
a sample size which is only about 1\% of the US population.
Thus, surprisingly, 
over the past two decades census data about marital status by age
have become less and less accurate. Incidentally,
it can be observed that
the CPR data are {\it systematically}
below the census data which shows that the differences cannot solely be
explained as being due to random sampling errors; there must also
be a non-sampling error component.\qL
{\it Sources:
Census 1980: Table 264 in the following 
census publication volume: US summary, Ch. D, Section A
(available on the website of the US Census Bureau);
CPR 1980: Series P-20, No 365, survey of March 1980 (issued in
October 1981); Census 2000: Table PCT007 available
on the FactFinder website of the US Census Bureau; CPR 2000: Series
P-20, No 537 (issued in June 2001).
ACS: Table B12002 available on the FactFinder website of the US Census
Bureau; ASES 2010: ``America's Families and Living Arrangements,
Supplement'', 2010. Many thanks to Dr. Rose Kreider from the US Census
Bureau for her help.}
\vskip 2mm
\hrule
\vskip 0.7mm
\hrule
\end{table}
%%-----------------------------------------------

\qA{Discussion of computational methods for estimating populations}

Before we close this section about population estimates
an additional observation is in order concerning 
computational methods.
\qpar

At first sight it might seem that in the intervals 
between census years it is easy to {\it compute}
the sizes of age-groups. Indeed,
based on the numbers of deaths, marriages, divorces 
in each age-group, one should be able to predict
the sizes of relevant age-groups.
Such a procedure which would permit to follow each age group
year after year until the next census 
may work in some countries, but in the United States
it does not. There are three main obstacles.
\qee{1} One does not know the flows of illegal immigrants.
Although this difficulty exists in all countries it is
more or less serious depending on the magnitude of 
illegal immigration.
\qee{2} The annual data about marriages and divorces are
known to be fairly incomplete in some US states. 
Until 1996 total divorces were reported by the Federal
Government. Subsequently, it ceased to publish national
divorce data.
\qee{3} Deaths which occur overseas are not included in the
death numbers published by the US Census Bureau. In other
words, the deaths of US soldiers in Europe, Korea, Vietnam,
Afghanistan or Iraq were not included in annual death statistics.
\qpar
It is true that fatality data are published by the Pentagon.
However, such data are incomplete in two respects.
\qbu  Firstly, the Department of Defense does not publish official
data for the fatalities among  
civilian contractors working for the armed forces.
Whereas, during the Vietnam War the proportion of military
personnel to civilian personnel was $ 6:1 $, during the
occupation of Iraq it was almost $ 1:1 $ (Flounders 2009).
In addition to
the personnel under contract there are also persons
who are not considered as contractors. For instance, one
can mention
news correspondents, businessmen, embassy personnel,
Peace Corps affiliates, members of the 
Young Men's Christian Association (YMCA), and so on.
When occurring abroad the deaths of such persons will
be basically unrecorded. Young age groups will be particularly
affected by such omissions.
\qbu Prior to 1980 the US Department of Defense
did not publish worldwide fatality data.
This point was made very clearly in 1993 when 
a data revision was announced by the Pentagon.
Previously it had been said that 54,246 soldiers had died
in the Korean War. According to the revision, there had been
36,516 deaths in Korea and  17,730 worldwide outside of the
war theater. For the Vietnam War, published data tell
us that there were 58,193 fatalities in the war theater but
no data are available for the military fatalities outside
the war theater.
\qpar
In terms of magnitude, the total of omitted overseas deaths
is certainly much smaller than the population of
illegal immigrants. 
\qpar

In spite of these difficulties, computational methods
are commonly used. For instance, in France population numbers by
marital status and age were computed for every 
year from 1901 to 1993 (Daguet 1995). The main
problem with such estimates is that it is impossible to
control their quality. Usually, in such calculations
one needs to make some assumptions, for instance that the death rate
for an age group truly reflects the change that occurred in this
age group. If for some reason (e.g. omitted overseas deaths)
these assumptions are not correct then the result of the calculation
will be biased. This is a non-sampling error which will not
be removed by taking averages over several years (at least
if the bias persists). The worst aspect is that researchers
who use such calculated figures have
no way whatsoever to control whether they are correct or not.
\qpar
On the contrary, for population estimates based
on samples, the resulting statistical uncertainty is well known
and in addition it can be reduced by averaging over several years.
Most of the data shown in this paper have been obtained in
that way. The only exceptions are Fig. 1,2,7.

\qI{Phase 1: Mid-age part of the Farr-Bertillon distribution}

Table 3b shows that, except for middle age-groups, the
accuracy provided by the ``Current Population Reports''
is fairly low especially for widowed persons.
Another concern is the existence of a systematic non-sampling
error component. In fact, we do not really know what population 
estimates were used to compute the death rates given
in the ``National Vital Statistics Reports'' that were
used in Fig. 2. The technical notes of the reports say only that
``the populations used to calculate death rates were
produced under a collaborative arrangement with the
US Census Bureau''. At least, this sentence suggests that
the population estimates were not merely drawn from
the CPR.  Probably the  CPR
were used as a starting point and, in some (unspecified)
way were corrected for small age-groups. Here the
omission of the
youngest age groups is a cautious and sensible step.

%%-----------------------------------------------
%% MAPM#MARITAL2 -> 15 ANNEES (1996-2010)
\begin{figure}[htb]
\vskip -5mm
\centerline{\psfig{width=11cm,figure=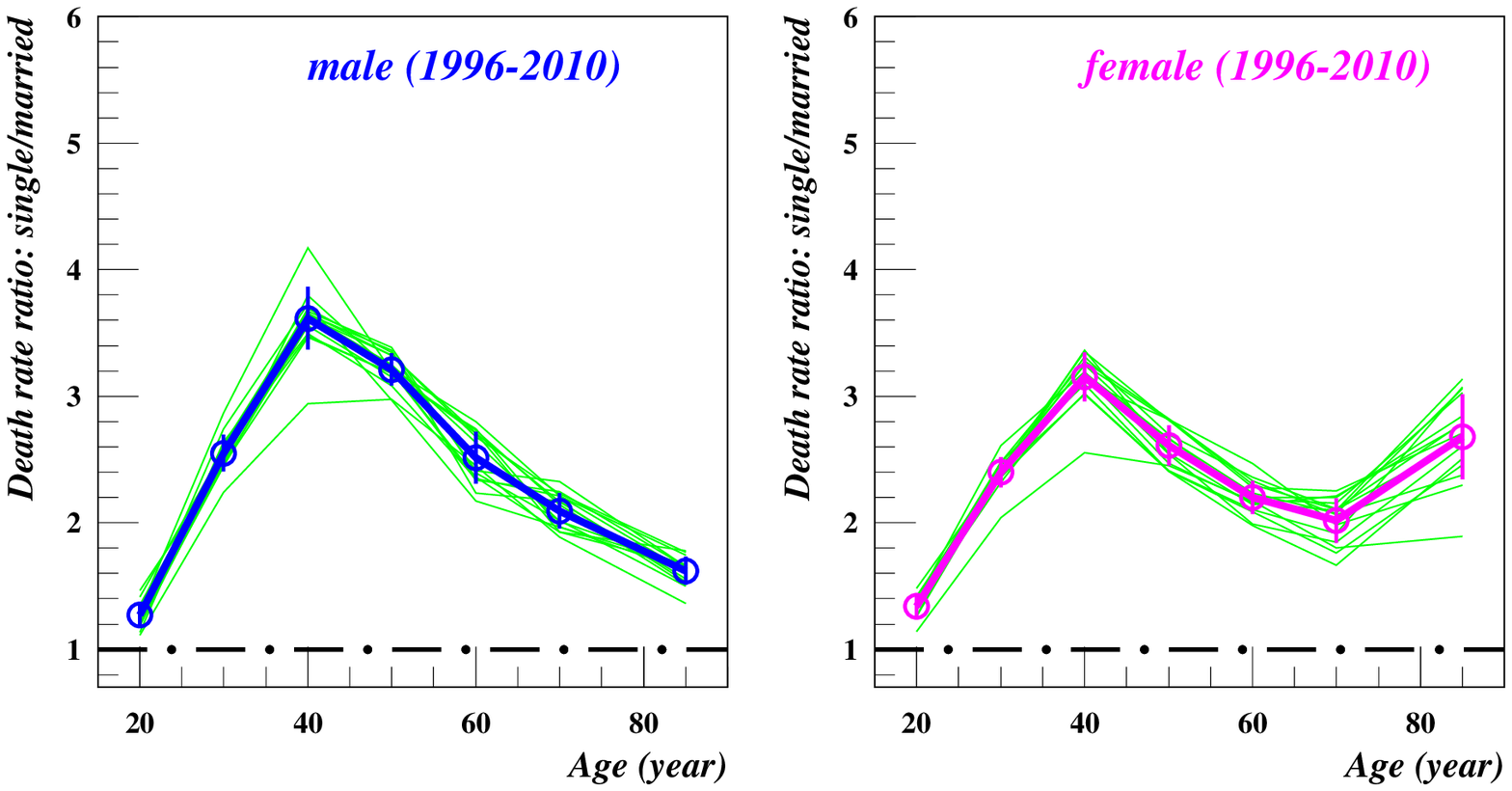}}
\vskip -5mm
\centerline{\psfig{width=11cm,figure=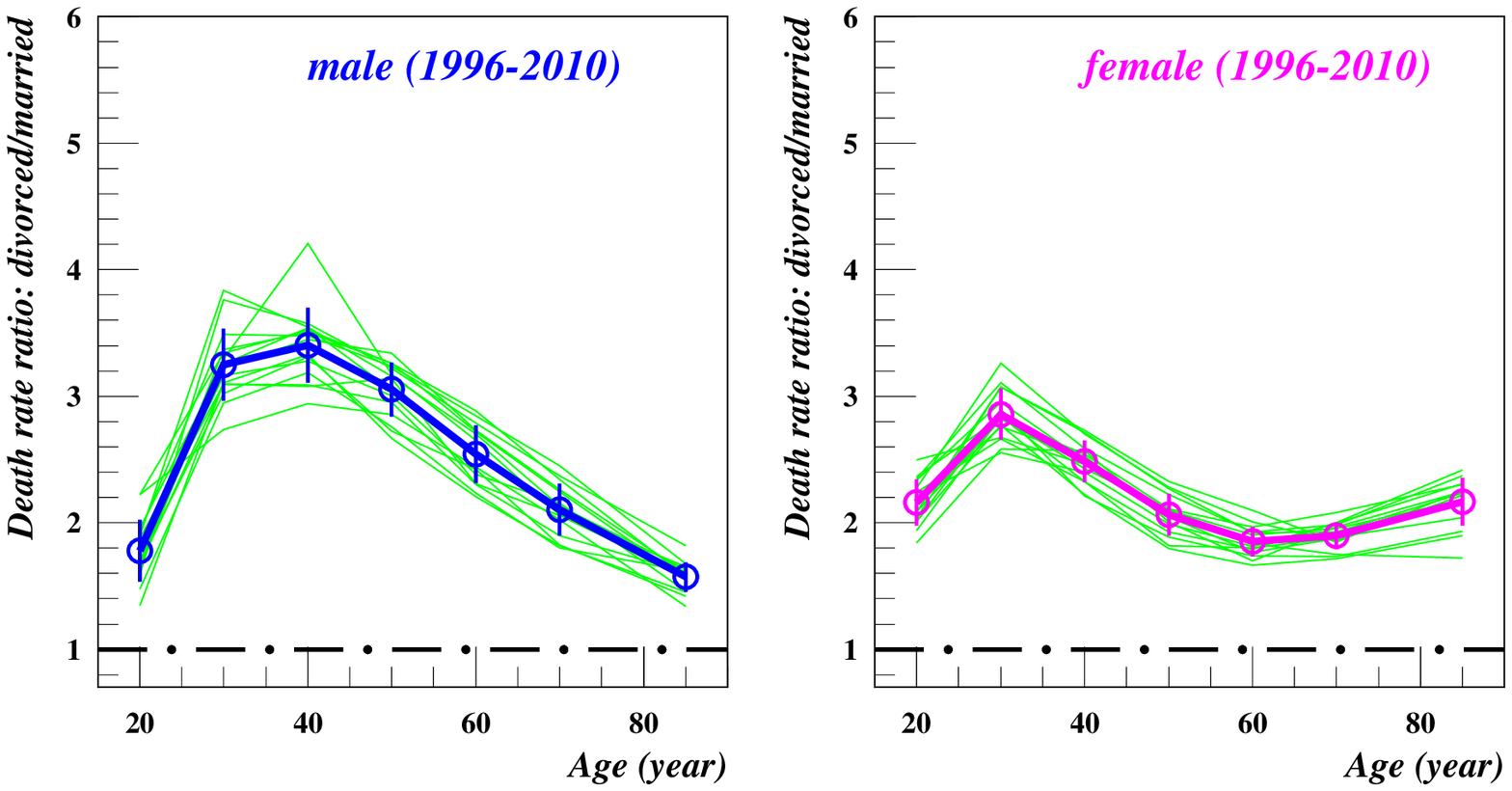}}
\vskip -5mm
\centerline{\psfig{width=11cm,figure=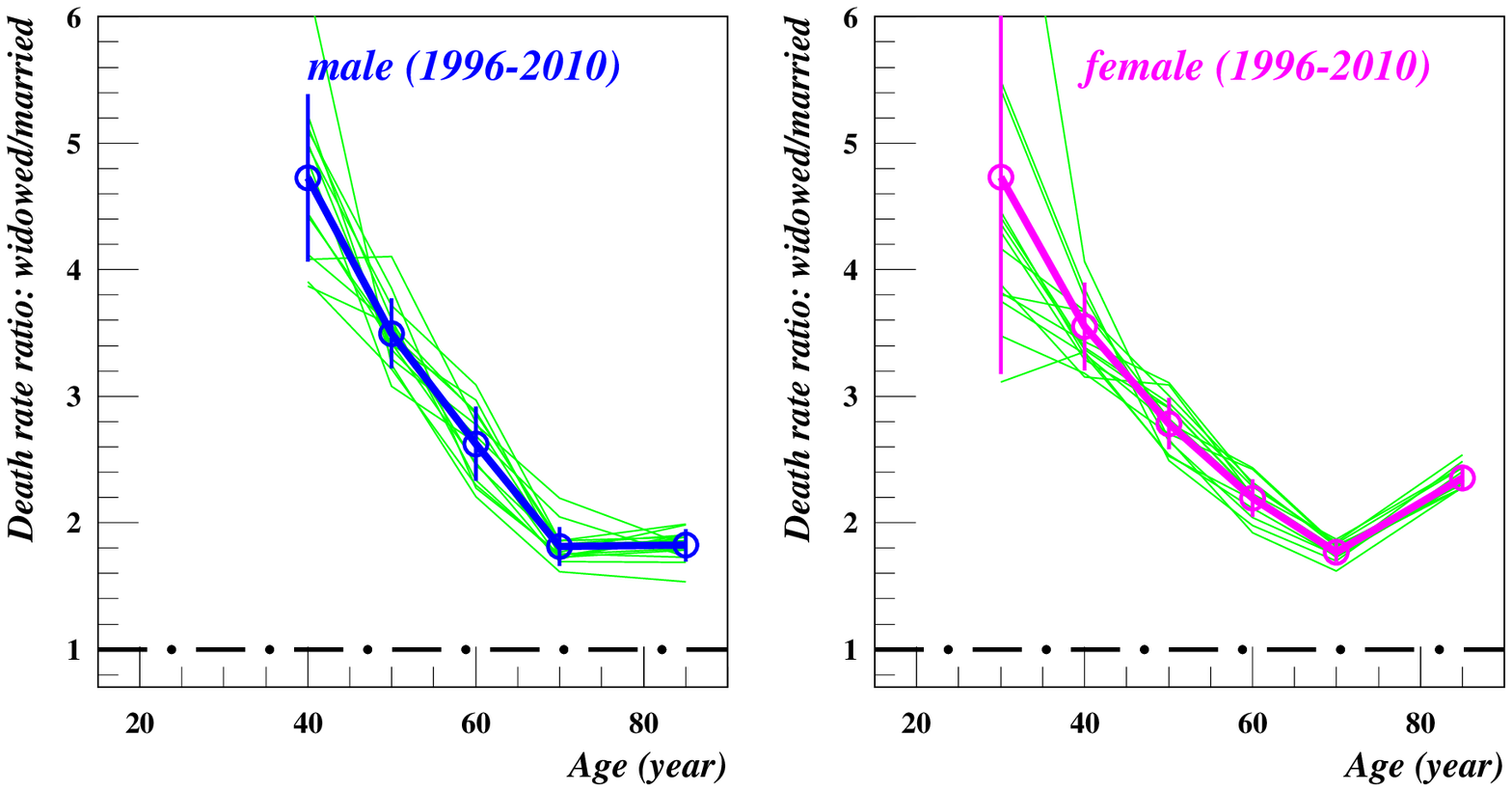}}
\qleg{Fig. 2a,b,c: Death rate ratio according to marital status
in the United States. From top to bottom: single/married,
divorced/married, widowed/married.}
{The thin (green lines) are the yearly curves for the 15 successive
years. The thick lines show averages over the 15 years. 
During this period
the ratios did not display any trend, there were only random
fluctuations.
There are 7 age groups ranging from $ 14-24 $ to $ 64-75,>75 $
but in the source no data are given for the youngest age groups
of the w/m case. In this graph as well as in all subsequent graphs
the length of the error bars is $ \pm 1.96\sigma $ 
(where $ \sigma $ is the standard deviation of the
average) which corresponds 
to a probability confidence level of 0.95.
There are no 
data points for small age groups (particularly for
young widowers/widows) because, apart from census years, 
there are no reliable estimates for the size of the corresponding
populations.}
{Source: National Vital Statistics Reports.
Deaths: Final Data. Successive years from 1996 to 2010. 
National Center for Health Statistics. The publication gives
the death numbers and the rates. How were these rates computed? 
The ``Technical notes'' attached to the table do not
give any specific information.  They say only that the
population data ``were 
produced under a collaborative arrangement with the
US Census Bureau''.}
\end{figure}
%-------------------------------------------------

\qI{Phase 2: The young widower effect}

In this section we present three sets of results.
\qbu The first series of graphs (Fig. 3)
is based on the censuses of 1940, 1950 and 1960.
For these data the rates were computed by the US Department
of Health and published in the data compilation done
by Robert Grove and Alice Hetzel (1968). 
\qbu The second series of graphs (Fig. 4) is based on
the censuses of 1980, 1990 and 2000. Actually 
in the census of 2000, marital status data were asked only 
on the so-called  ``long-form'' which was distributed to
20\% of the households. 

%%-----------------------------------------------
%% MAPM#MARITAL5 -> 1940, 1950, 1960 (Grove and Hetzel)
\begin{figure}[htb]
\vskip -5mm
\centerline{\psfig{width=13cm,figure=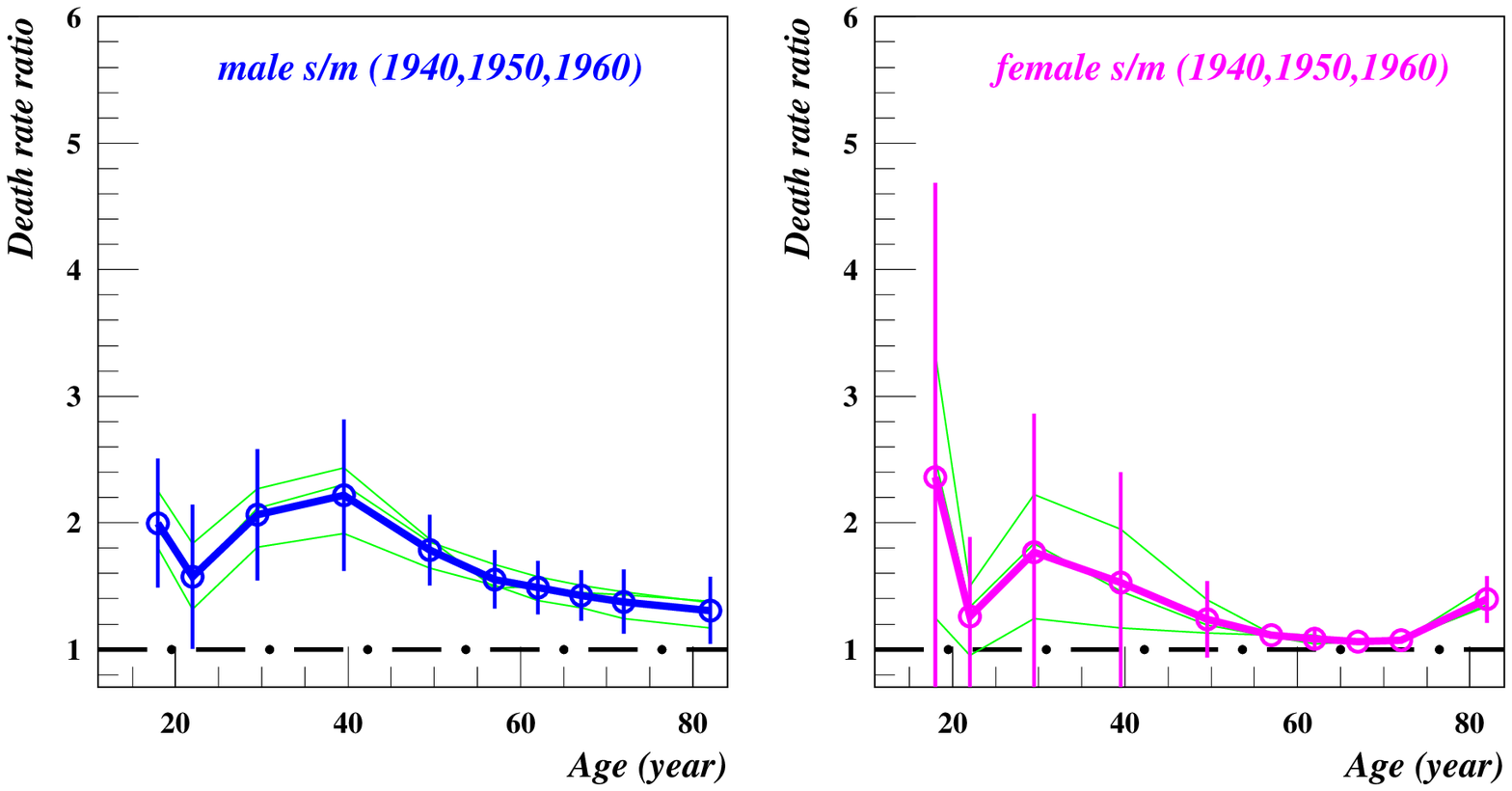}}
\vskip -5mm
\centerline{\psfig{width=13cm,figure=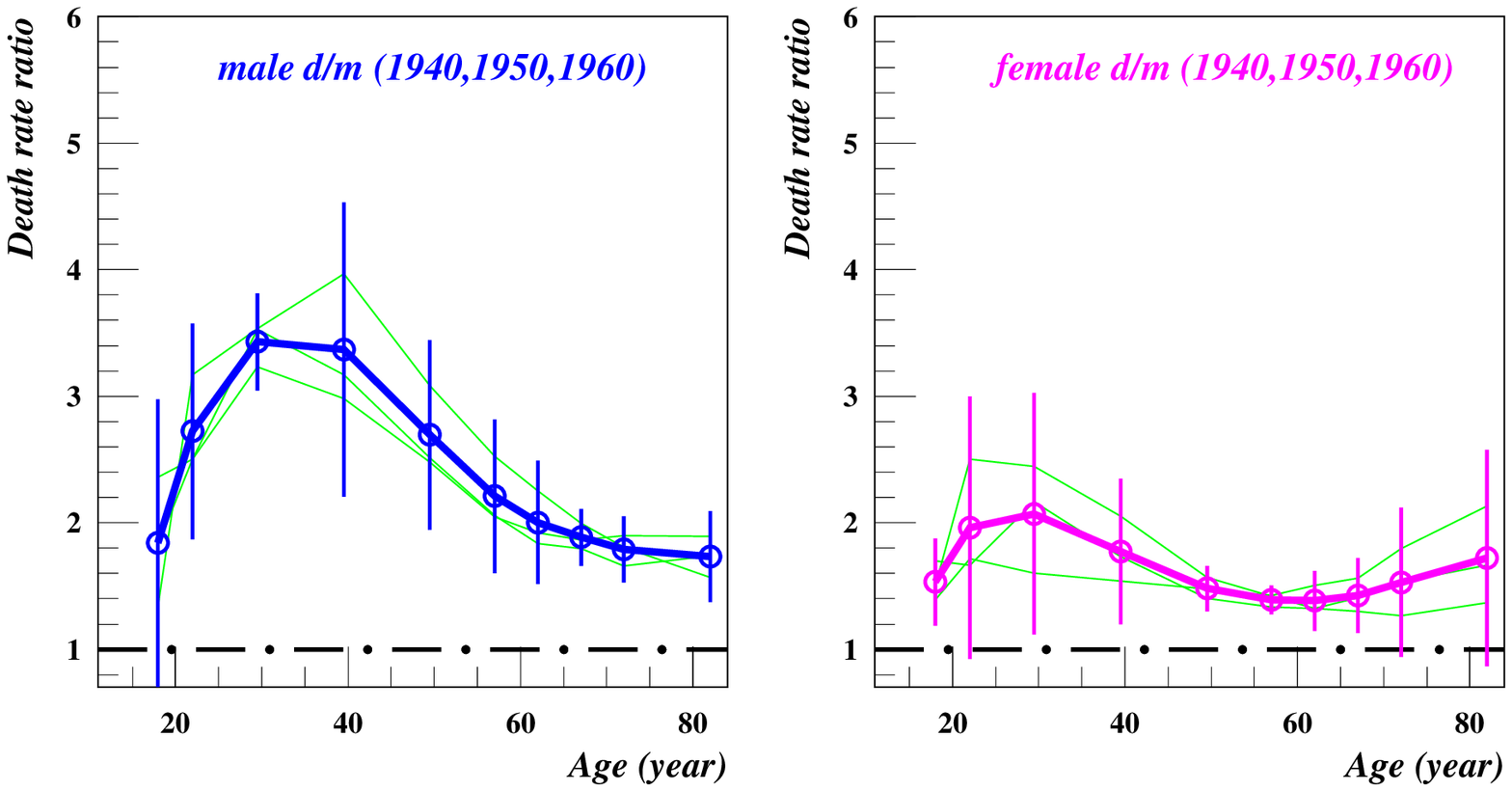}}
\vskip -5mm
\centerline{\psfig{width=13cm,figure=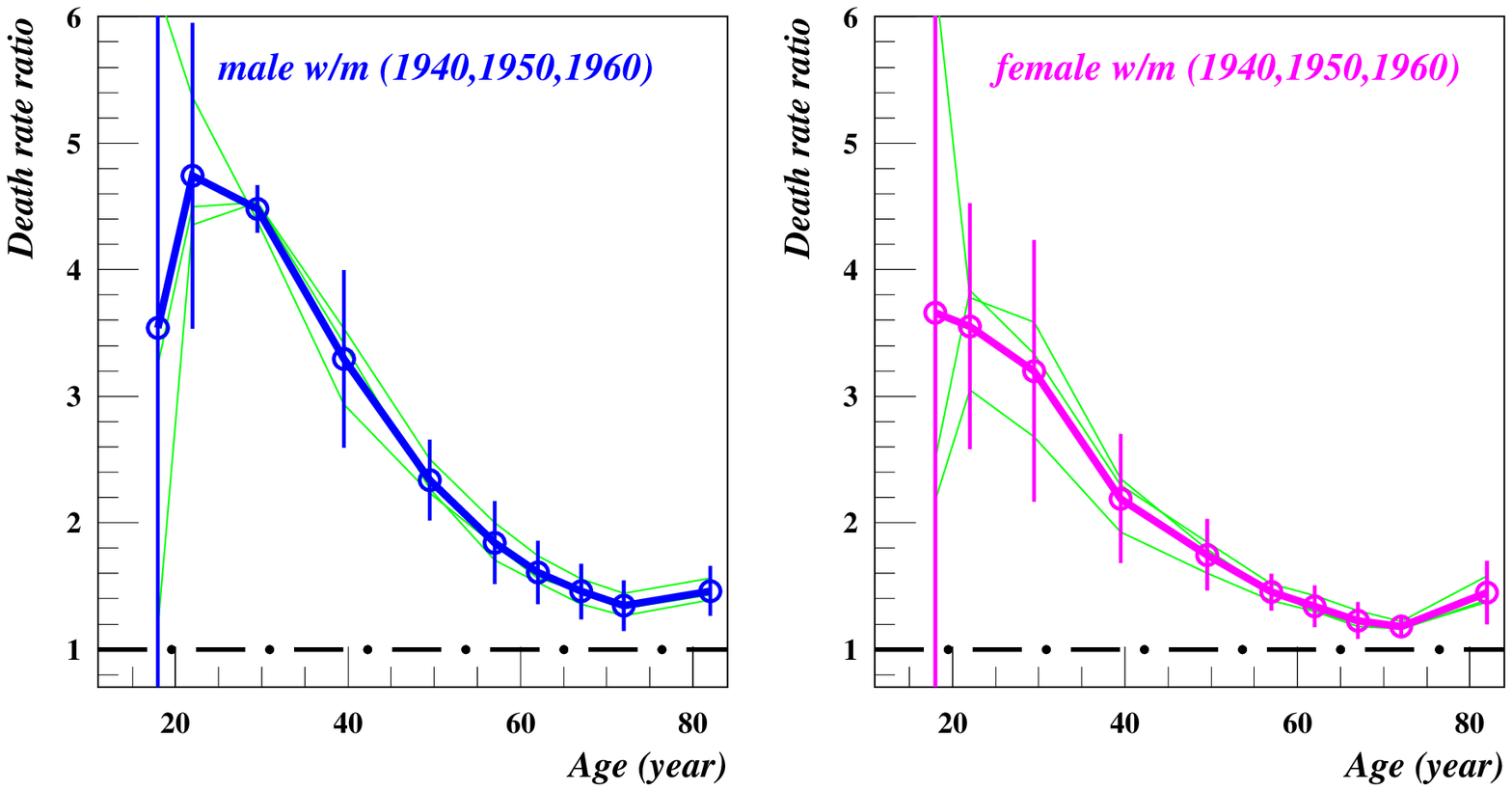}}
\qleg{Fig. 3a,b,c: Death rate ratio with respect to
the death rate of married persons:
United States, 1940, 1950, 1960. 
From top to bottom: single/married,
divorced/married, widowed/married.}
{The thin (and green) curves are for each of the 3 years while
the thick lines with the round dots show their average.
There are 10 age groups: $ <20,20-24,25-34,35-44,45-54,55-50,60-64,
65-69,70,74,>75 $
As expected, the error bars become fairly large for small
age groups, particularly for young widowers and widows.} 
{Source: Grove and Hetzel (1968, p.334). The publication
gives directly the rates.}
\end{figure}
%-------------------------------------------------

%%-----------------------------------------------
%% MAPM#MARITAL6 -> 1980, 1990, 2000 (WIDOWED)
\begin{figure}[htb]

\centerline{\psfig{width=13cm,figure=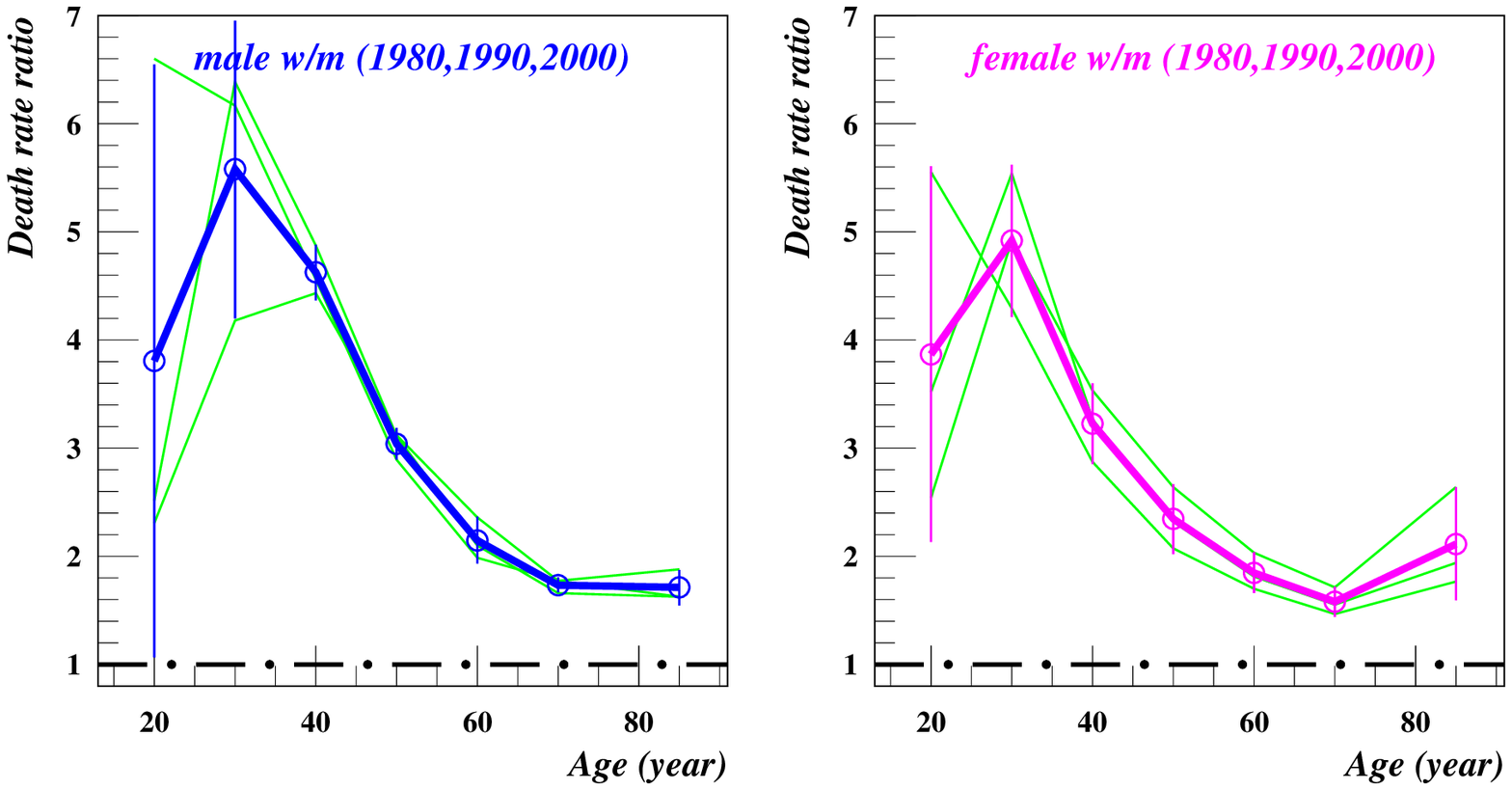}}
\qleg{Fig. 4: Death rate ratio for widowed persons (i.e. 
their death rate
divided by the death rate of married persons),
United States, 1980, 1990, 2000.}
{The thin (and green) curves are for each of the 3 years while
the thick lines with the round dots show their average.
The age groups are the same as in Fig. 2.
As expected, the error bars become fairly large 
for young widowers and widows.\qL} 
{Sources: Census population data:
1980: Vol. PC80-1-D1-A (Table 264, p. 1-67);
1990: General population characteristics, Vol. CP-1-1, (Table 34, p. 45);
2000: ``American FactFinder'' website of the US Census Bureau, Table
PCT007. \qL
Mortality data: 
1980: Vital Statistics of the United States 1980, Vol. 2, part A, 
Table 1-31 (p. 315); 
1990: Vital Statistics of the United States 1990, Vol. 2, part A, 
Table 1-34 (p. 387);
2000: National Vital Statistics Report, Vol. 50, No 15, 16 Sep 2002.
}
\end{figure}
%-------------------------------------------------

\qbu In Fig 3 and 4 the error bars for widowed persons in the
young age groups remain very large. In an attempt to
reduce them, we use data for 6 successive years.
These data are based on the ``American Community Survey''
which is answered by about 1\% of the households.
Thanks to the 6-year interval, the error bars are notably reduced.  

%%-----------------------------------------------
%% MAPM#MARITAL7 -> 6 ANNEES 2005-2010 PAR ACS
\begin{figure}[htb]
\centerline{\psfig{width=13cm,figure=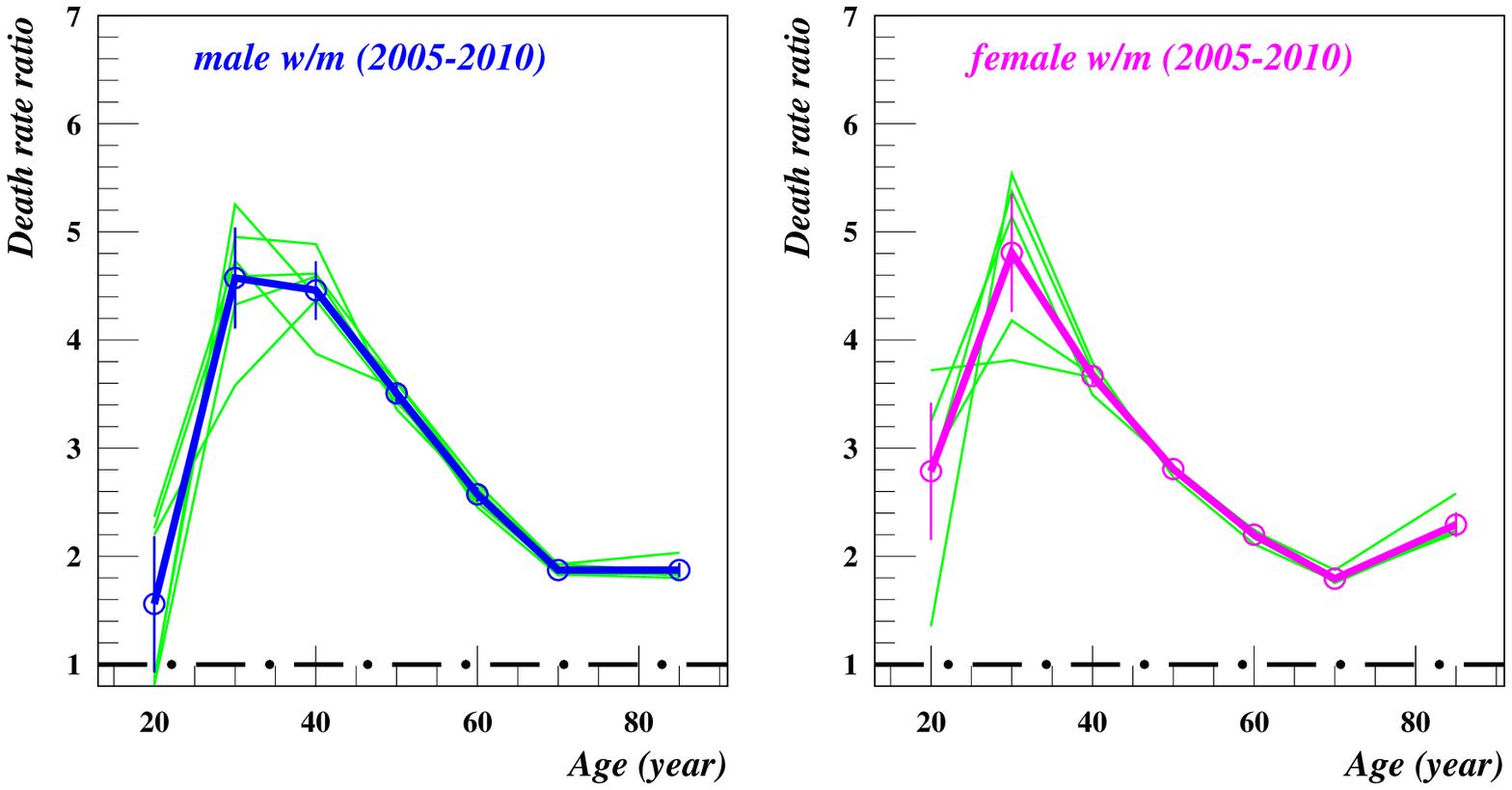}}
\qleg{Fig. 5: Death rate ratio of widowed persons,
United States, 2005-2010.}
{The thin (and green) curves are for the 6 successive years while
the thick lines with the round dots show their average.
The age groups  are the same as in Fig. 2.
The length of the error bars is $ \pm 1.96\sigma $ which corresponds 
to a probability confidence level of 0.95.
The data used in this graph for the populations are not
census data but are based on samples of about 2.5 million
respondents.
This ``experiment'' confirms 
the existence of a dip for the youngest age group $ 15-24 $.\qL} 
{Sources: Populations: Starting in
2005 data by marital status and age
are provided by the ``American Community Survey'' 
and are made available 
on the ``American FactFinder'' website set up by the US
Census Bureau (Table B12002).
Mortality data: National Vital Statistics Reports entitled
``Deaths: Final Data''.
In 2015 the most recent year available was 2010.}
\end{figure}
%-------------------------------------------------

\qA{Error bars}

As stated in the caption of Fig. 2, the length of the
error bars are $\pm 1.96\sigma(\hbox{average}) $.
The standard deviation of the average was computed by dividing
the standard deviation of $ n $ annual curves $ Y_j,\ j=1,\ldots,n $
by the standard
$ 1/\sqrt{n} $ factor. However, this factor is correct only
when the $ Y_j $ are not correlated. While there is indeed
a low correlation for young age groups, for older age groups
there is an average correlation of $ r_m=0.90 $. For these
data points the factor $ 1/\sqrt{n} $ should be replaced
by $ f=\sqrt{1+(n-1)r_m}/\sqrt{n} $ (Roehner 2007, p. 45). 
With $ n=6 $ and $ r_m=0.90 $
the factor $ f $ is almost equal to 1. In other words, except
for young age groups, the error bars shown in the graph 
underestimate the actual confidence intervals. On the other
hand, using the factor $ f $ everywhere would result
in overestimating the confidence interval for young age groups,
the only ones which really matter in this respect.

\qA{What is the influence of cohabitation and separated couples?}

In recent decades the traditional picture of family life has
become more complicated due to the following trends.
\qee{1} In 1960, 72\% of all American adults were married; in 2012 just 
50\% were. 
\qee{2} During the same time interval, the number of 
cohabiting non-married couples of opposite sex jumped from
1.1\% to 11\%. Note that because different states
may not have the same definition of cohabitation the last
percentage may be subject to an error margin of about
$ \pm 10\% $. 
\qee{3} Finally around 2005, in about 8\% of married couples
one of the spouses was not present. In 1980 this proportion
was about 6\%.
\qpar

Needless to say, such trends are by no means
special to the United States; they are shared by many other western
countries. However, the trends are perhaps more surprising in the
United States because traditionally this country has put a strong
social emphasis on family life. In this respect it can be recalled
that in 2014
three states, Mississippi, Michigan, and Florida, still had laws
against cohabitation by opposite-sex couples.
\qpar

How do the previous trends affect the interpretation of our results?
\qpar
We will successively
consider the effects of cohabitation and separation

\qA{Implication of cohabitation}

In itself point 1 will not affect our results but in fact
it is strongly connected with the second point: those people who do
not get married are doing so because they are living together without
being married. \qL
To make the argument clearer let us make the following
simple assumptions. We assume that the {\it real} death rates
in the married, single and widowed classes are 1, 2 and 3 per 1,000.
In addition we assume that 50\% of the persons registered
as single or as widowed are in fact cohabiting with a partner.
For the sake of simplicity we assume that there are 2,000 single
and 2,000 widowed persons.
Moreover we assume that the measured death rate of married
people is not modified and is equal to the real death rate.
This makes sense as long as the number
of cohabiting people remains small compared to married people.
\qpar
Under these assumptions, what
are the death rates, $ d_{me} $, and death rate ratios, $ r_{me} $,
that will be measured and how do they compare to the real
death rate ratios $ r_{re} $?
\qbu {\bf Death rate ratio of single persons}
$$ d_{me}(s)=(2+1)/2=1.5\quad  \rightarrow \quad  r_{me}(s)=1.5/1=1.5,\quad
\hbox{vs. } r_{re}(s)=2/1=2 $$
\vskip -8mm
\qbu {\bf Death rate ratio of widowed persons}
$$ d_{me}(w)=(3+1)/2=2.0\quad  \rightarrow \quad  r_{me}(w)=2.0/1=2.0,\quad
\hbox{vs. } r_{re}(w)=3/1=3 $$

In other words, due to cohabitation 
our measurements will underestimate the
actual death rate ratios of single, and widowed persons. The same
argument applies of course to divorced persons.

\qA{Implication of separation}
In the present argument we suppose that there is no cohabitation
which means that the death rates of single and widowed persons
are correct. In addition, in the same way as above, we assume that
50\% of the married persons are in fact separated. With
the same real death rates as above what will be the measured
death rate ratios? Whereas previously, the numerators of the death
rate ratios were affected, this time the denominators are affected.
\qbu {\bf Death rate ratio of single persons}
$$ d_{me}(m)=(1+2)/2=1.5\quad  \rightarrow \quad  r_{me}(s)=2/1.5=1.33,\quad
\hbox{vs. } r_{re}(s)=2/1=2 $$ 
\vskip -8mm
\qbu {\bf Death rate ratio of widowed persons}
$$ d_{me}(m)=(1+2)/2=1.5\quad  \rightarrow \quad  r_{me}(w)=3.0/1.5=2.0,\quad
\hbox{vs. } r_{re}(w)=3/1=3 $$

In other words, separation will also make 
our measurements  to underestimate the
actual death rate ratios.
\qpar
Because the two effects go in the same direction, their combination
should also result in underestimating the real death rate ratios.
An additional conclusion is that if we see a weakening of the
Farr-Bertillon effect over the coming decades it may well
be a statistical artifact due to persistent cohabitation and separation
trends.

\qI{Death ratios by marital status and age for selected causes}

In this section we consider death rate ratios according to
selected causes of death. 
Previously we have seen that the death ratios for
never-married and divorced persons are somewhat similar.
Therefore, 
the present investigation will be restricted to
never-married and widowed persons.
\qpar

%%-----------------------------------------------
%% MAPM#MARITAL8 -> PAR CAUSE DE DECES, 1980 + 1990 
\begin{figure}[htb]
\vskip -5mm
\centerline{\psfig{width=13cm,figure=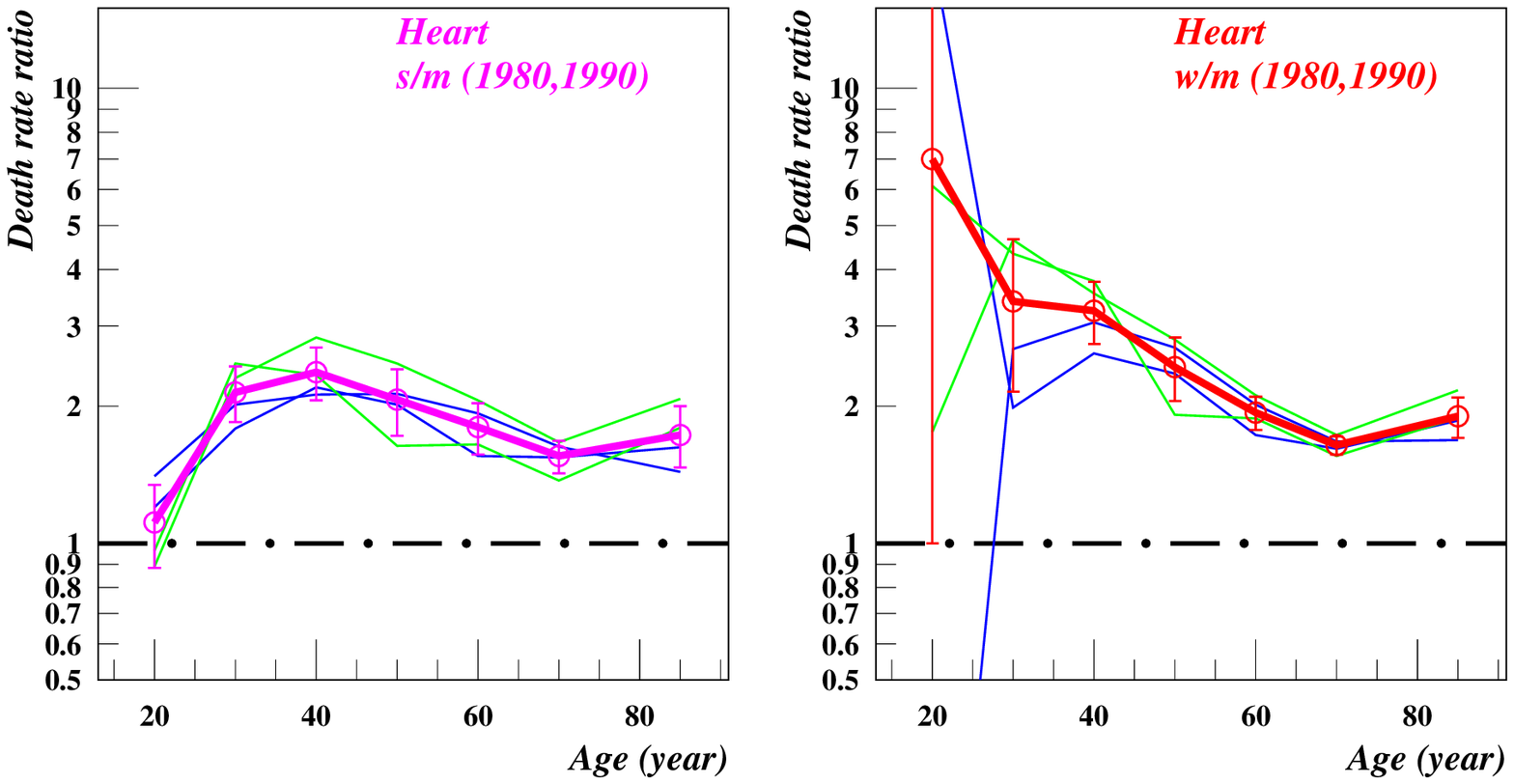}}
\vskip -5mm
\centerline{\psfig{width=13cm,figure=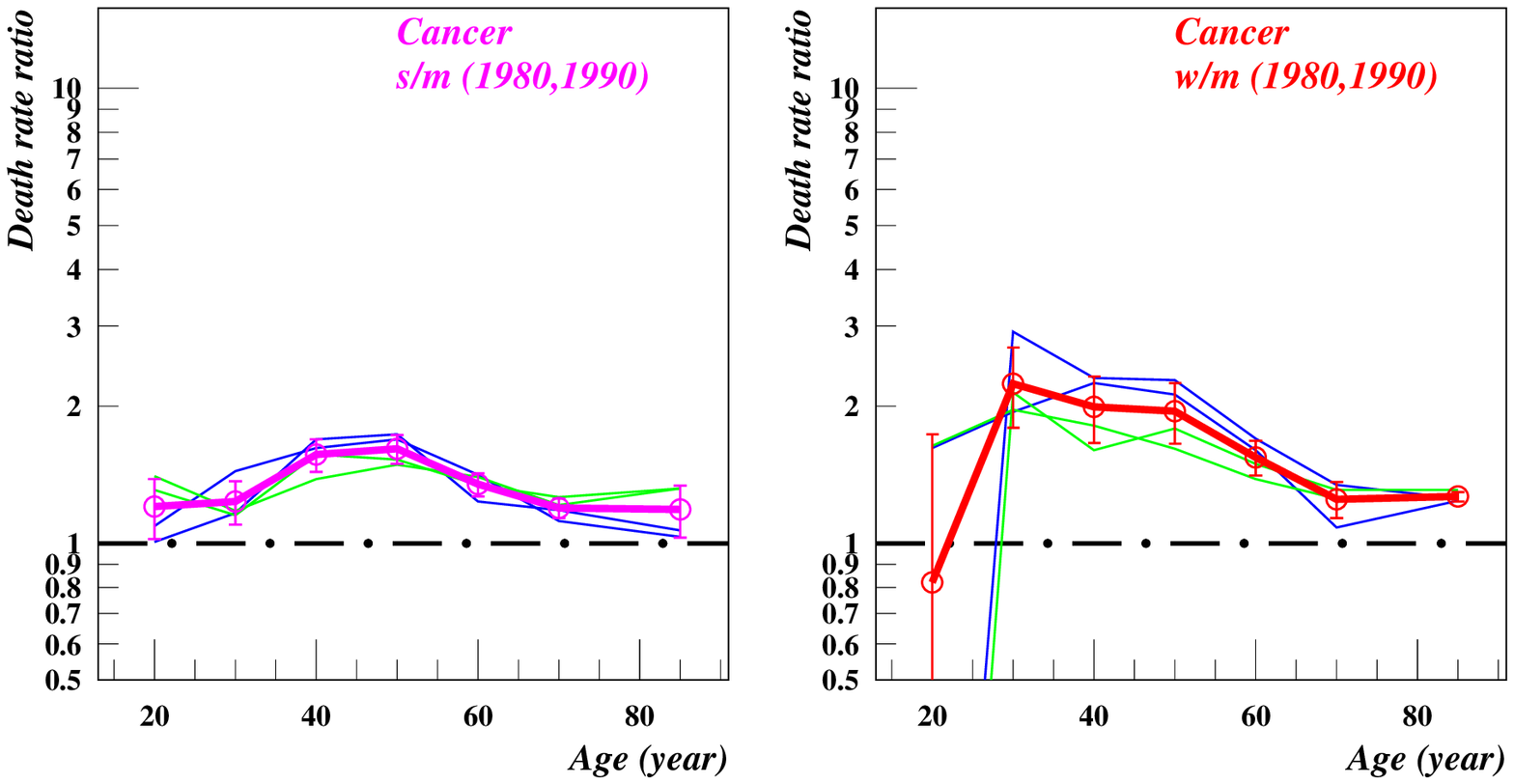}}
\vskip -5mm
\centerline{\psfig{width=13cm,figure=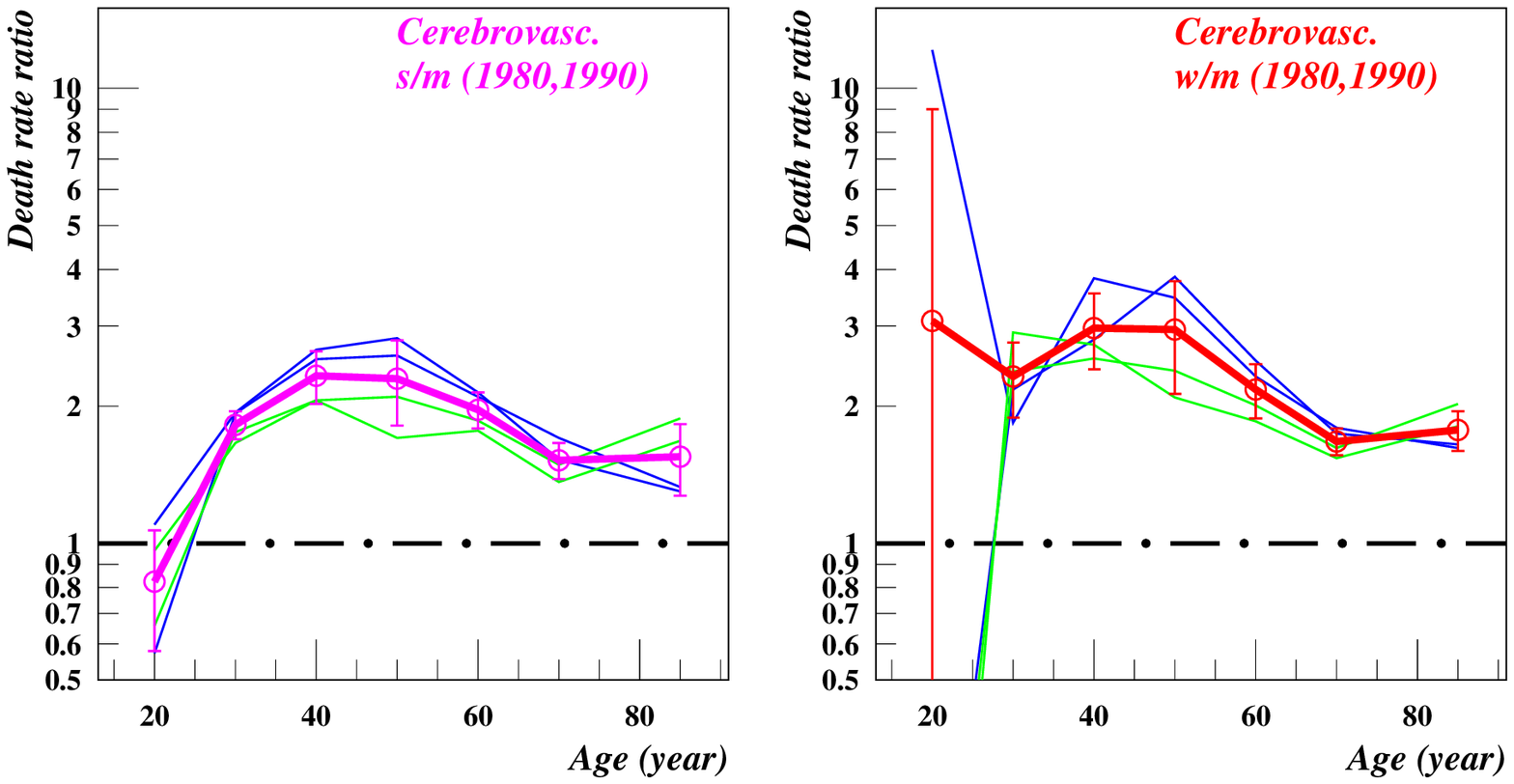}}
\qleg{Fig. 6a,b,c: Death rate ratio for different causes of death,
United States.}
{The $ s/m $ graphs on the left-hand side are for single/married,
whereas the $ w/m $ graphs are for widowed/married.
The graphs show 1980 and 1990 data for males
(thin blue lines) and for females (thin green lines).
The thick lines represent the averages of the 4 series. 
The age groups are the same as in Fig. 2.}
{Sources: 1980: Vital Statistics of the United States, 1980, Vol. 2, Part
A, table 1-31, p. 315-324.
1990: Vital Statistics of the United States, 1990, Vol. 2, Part
A, table 1-34, p. 387-400.
In order to compute the rates we used census population data which
is why the analysis is restricted to census years.}
\end{figure}
%-------------------------------------------------

%%-----------------------------------------------
%% MAPM#MARITAL8
\begin{figure}[htb]
\vskip -5mm
\centerline{\psfig{width=15cm,figure=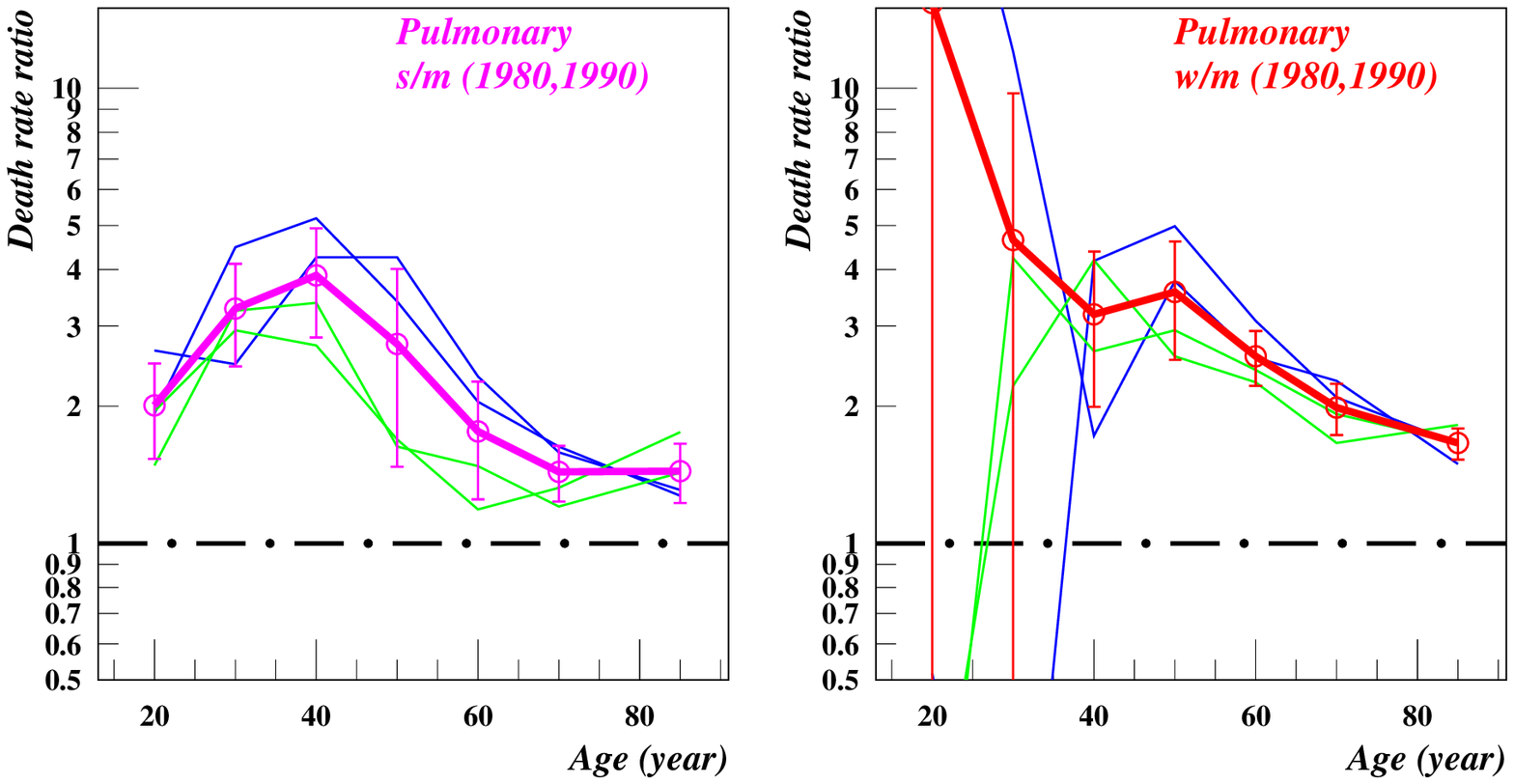}}
\vskip -5mm
\centerline{\psfig{width=15cm,figure=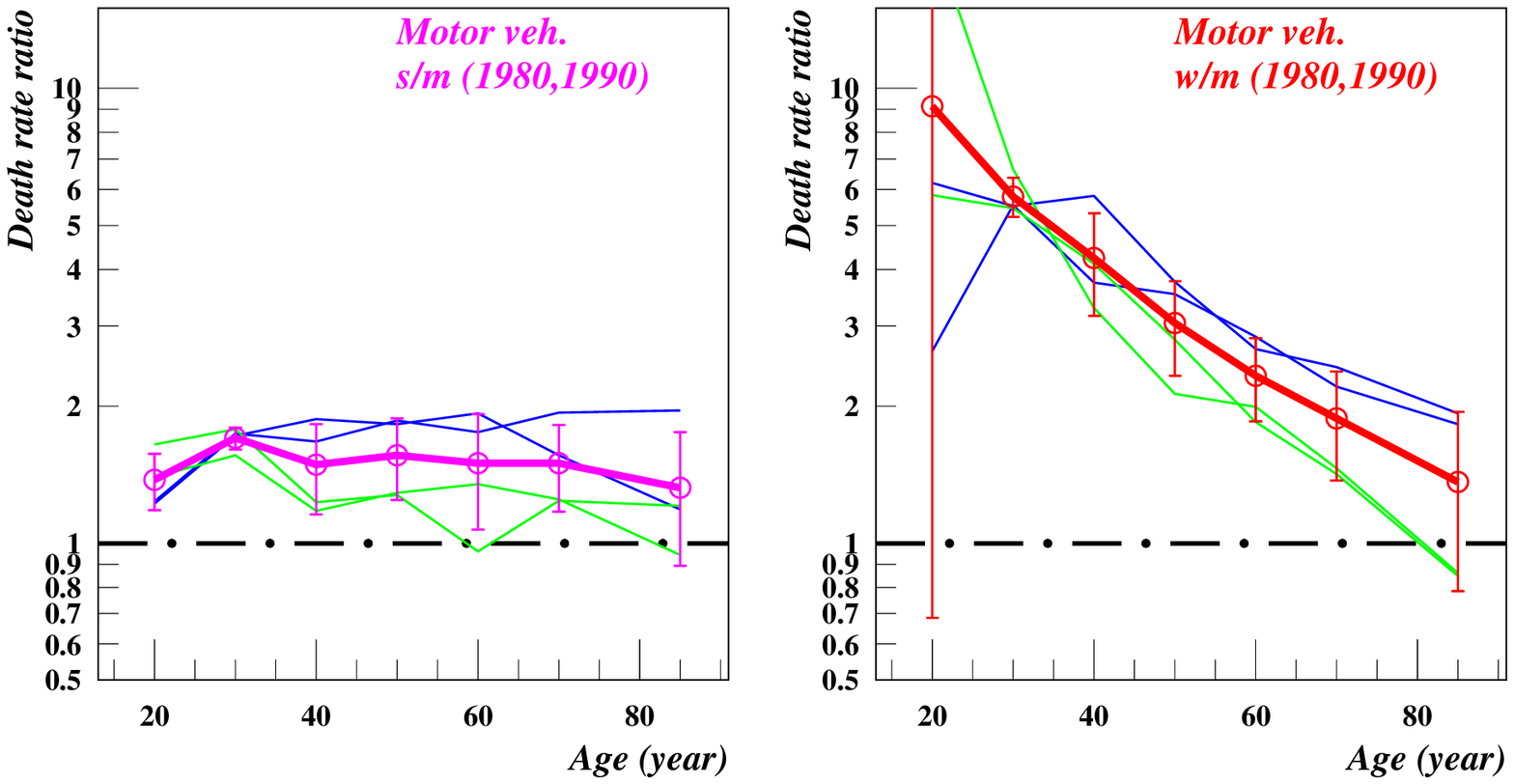}}
\vskip -5mm
\centerline{\psfig{width=15cm,figure=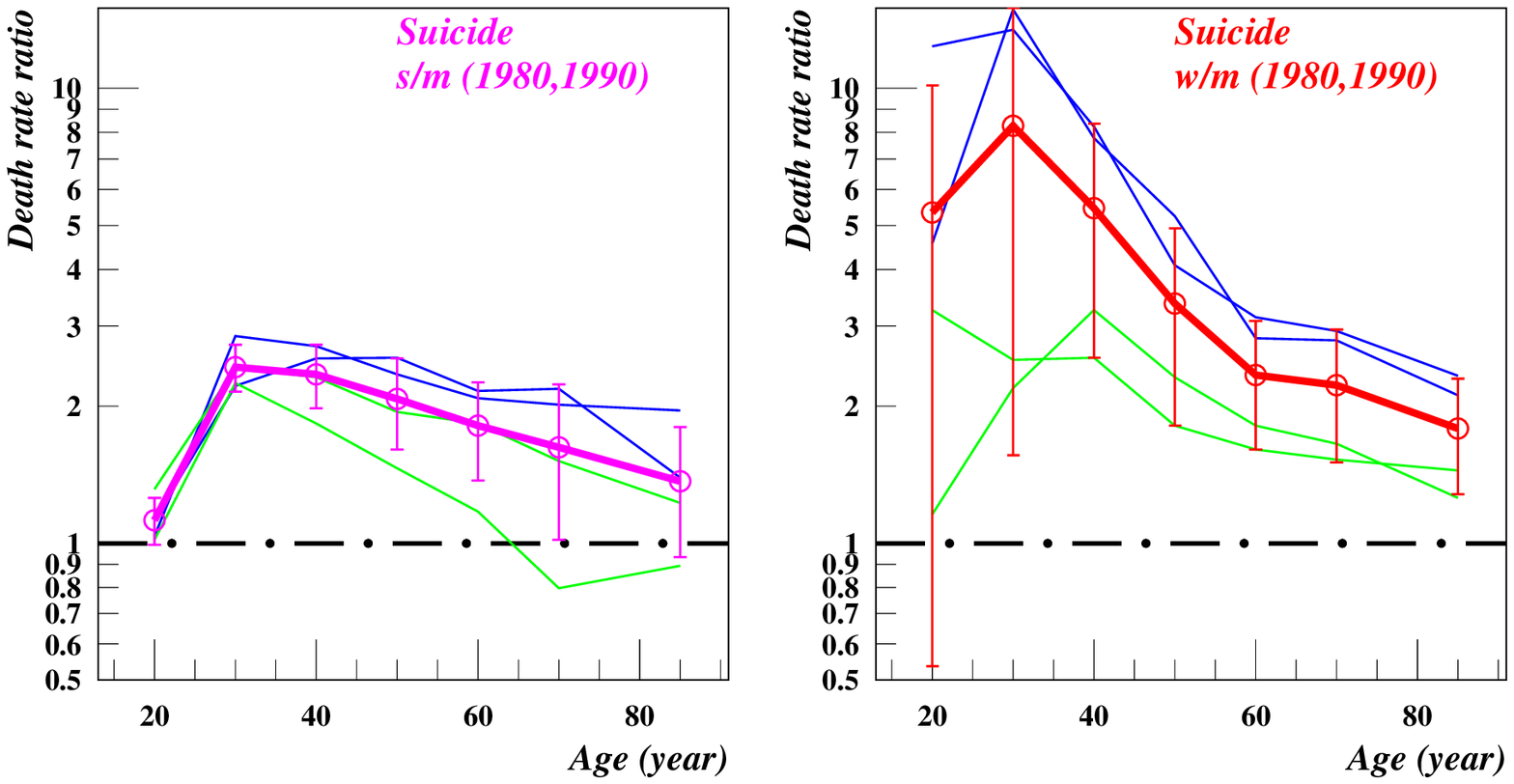}}
\qleg{Fig. 6 d,e,f: Death rate ratio for different causes of death,
United States (continued).}
{The comments made in Fig. 6 a,b,c apply as well to the
present graphs.}
{Sources: Same as for Fig. 6 a,b,c}
\end{figure}
%-------------------------------------------------

The main difficulty with data by cause of death is the fact
that the death numbers are fairly small which creates large
fluctuations. In an attempt to smooth them out 
as far as possible we lumped together not only successive years
(as was already done in previous graphs) 
but also the two genders.\qL
Incidentally, it can be noted that death numbers by cause of death,
marital status and age exist from 1979 to 1993, but only 
1980 and 1990 can be used because for the populations
we must rely on census data.
\qpar

How well do the 6 selected causes of death represent all causes?
For married persons, 
in the age-group 15-24 the cumulative death rates
(per 100,000) 
of the 6 selected causes total 51 (average of 1980 and 1990);
that is only slightly less
than the total for {\it all} causes which is 74.
\qpar

The graphs in Fig. 6  can help us to
better understand the origin of the
fluctuations observed in the left-hand side
of the curves for widowed persons.
For convenience we denote the first two
data points of these curves by $ r_1 $ and $ r_2 $.
In some graphs $ r_1 $ was higher
than $ r_2 $, whereas in others it was the opposite. Why?
In order to connect the aggregated curves of Fig. 4-5
with those
displayed in Fig. 6 we 
must first ask ourselves what are the leading causes of death.
In the age-group $ 15-24 $ the
main factors are motor vehicle accidents with a rate
of 31 per 100,000 (average rate for married persons in
1980 and 1990), followed
by suicide with a rate of 11 and cancer with a rate of 5.1.
Thus, in this age-group, 
motor vehicle accidents represent almost one half of the
total death rate. Despite this particular factor being
larger than all the others, it
has huge fluctuations. Indeed, the error bars for motor vehicle
accidents are much broader than for cancer or heart disease
for which death numbers are much smaller.
\qpar
In short, it appears that the uncertainty affecting the
youngest age groups is due to the very high volatility
assocated with motor vehicle death statistics.

\qA{Ranking of causes of death according to death ratios}

Because almost all death ratios documented in Fig. 6 are
larger than 1 it makes sense to consider averages
over all age groups. This will allow a ranking of the
causes of death according to their death ratios (Table 4).

%%-----------------------------------------------
\def\err#1#2{{\displaystyle #1}{\scriptstyle \pm #2}}

\begin{table}[htb]

\centerline{\bf Table 4: Ranking of causes of death according
to their average death ratio}

\vskip 3mm
\hrule
\vskip 0.7mm
\hrule
\vskip 1.2mm

\color{black} 
\small

$$ \matrix{
\hbox{\color{blue} Marital status} \hfill& & & & & &
\hbox{\color{blue} Average}\hfill\cr
\qtb
\hbox{} \hfill& & & & & & \hbox{\color{blue}(all 5 causes)}\hfill\cr
\noalign{\hrule}
\qth
\hbox{\bf \color{blue} Never married} \hfill& \hbox{}&\hbox{}
&\hbox{} 
&\hbox{}  &\hbox{} &\cr
\hbox{\bf } \hfill& \hbox{suicide}&\hbox{heart}
&\hbox{cerebrovasc.} 
&\hbox{motor veh.}  &\hbox{cancer} &\cr
\hbox{\quad Death ratio s/m} \hfill & \err{1.8}{0.4}&\err{1.8}{0.3} 
& \err{1.76}{ 0.4}&
\err{1.5}{0.1} & \err{1.3}{0.1}& \err{\color{blue}1.6}{0.1}\cr
\hbox{} \hfill& \hbox{}&\hbox{} &\hbox{} 
&\hbox{}  &\hbox{} &\cr
\hbox{\bf \color{blue} Widowed} \hfill& \hbox{}&\hbox{} &\hbox{} 
&\hbox{}  &\hbox{} &\cr
\hbox{\bf } \hfill& \hbox{suicide}&\hbox{motor veh.} &\hbox{heart} 
&\hbox{cerebrovasc.}  &\hbox{cancer} &\cr
\qtb
\hbox{\quad Death ratio w/m}\hfill & \err{4.1}{1.7}& \err{4.0}{2}&
\err{3.1}{1.4}&
\err{2.4}{0.7} & \err{1.6}{0.4}& \err{\color{blue} 3.0}{0.4}\cr
\noalign{\hrule}
}
$$
\vskip 2mm
Notes: The figures given in this table are averages over
the 7 age groups considered in Fig. 6. Overall the death ratio
for w/m is about twice the death ratio of s/m. However, 
with the exception of motor vehicle accidents, the
ranking is almost the same. The error bars are for a
probability confidence level of 0.95.
The ``pulmonary disease'' cause of death has not been
included in this ranking because it has very large fluctuations:
its coefficients of variation 
are $ 40\% $ for s/m and $ 102\% $ for w/m.
The 6 causes of death under consideration correspond to the following
code numbers in the 9th Revision of the International
Classification of Diseases of 1975: heart diseases: $ 390-398+404-429 $,
cancer: $ 140-208 $, cerebrovascular diseases: $ 430-438 $, 
pulmonary diseases: $ 490-496 $, motor vehicle accidents: $ E810-E825 $,
suicide: $ E950-E959 $.
\qL
{\it Sources: Same sources as for Fig. 6}
\vskip 2mm
\hrule
\vskip 0.7mm
\hrule
\end{table}
%%-----------------------------------------------

Because the w/m death ratios are based on smaller population
numbers than the s/m ratios they have higher volatility.
Nevertheless, the
fact that the ranking of causes of death is almost the same
in the two cases shows that overall the w/m ratios are
trustworthy.

\qA{Marital ties as a most effective drug}
Can marriage be considered as an effective multipurpose drug?
Yes and no.
\qpar
``No'' for a very obvious reason: it can only reduce the death rates 
of persons who are not already married. What proportion
do non-married persons (to be
distinguished from the never-married)
represent in the age-group $ 65-74 $?
In 2005 for instance, according of the data provided by 
the ``American Community Survey'' the non-married  were
$ 21\% $ for men and $ 43\% $ for women.
\qpar
The previous question can also be answered affirmatively because
for single or widowed persons, marriage
makes really a big difference. In this respect, one should remember
that in clinical test trials most pharmaceutical drugs, for instance
those against heart disease, provide at most a 20\% -- 30%
benefit (more details can be found in Roehner 2014).
On the contrary, Fig. 6 shows that except for cancer in old age,
the death rate is divided at least by a factor 1.5, which
represents a reduction of 33\%, and in many cases the reduction
is over 50\% (division by 2).

\qI{Suicide}

\qA{Why is suicide of special interest?}

Among the causes of death considered previously, suicide has
a special significance for (at least) four reasons.
\qbu Historically, since the mid 19th century,
the phenomenon of suicide arose considerable interest among
sociologists. The work of Emile Durkheim (1897) is probably the most
well known but there were many other studies, for instance
by Louis-Adolphe Bertillon and his son Jacques Bertillon.
\qbu 
Durkheim showed that persons with many family links
have smaller suicide rates. 
For married persons with respect to never-married or
widowed persons, this connection
was already well-known before Durkheim.
Although the influence of the number
of children had also been observed (particularly
by Bertillon, see above),
the relationship was not known accurately because of the
fact that death
certificates did not contain information about
the number of children. Durkheim was able to establish
the existence of a negative correlation between number
of children and suicide rates by
taking advantage of regional
differences in suicide rates on the one hand and in average
number of children by household on the other hand.
This observation (largely forgotten nowadays)
was a strong argument in favor of
Durkheim's thesis of a connection between the strength of 
family ties and low suicide rates.
\qbu In biology a phenomenon called apoptosis
is often referred to as ``cell suicide''.
Apoptosis occurs when cells no longer
receive ``stay alive'' signals from their
neighbors. More details
can be found in Raff (1998), in chapter 12 of Roehner (2007)
and in Wang et al. (2013).
This link with apoptosis gives at least a plausible
mechanism for the connection between suicide and interaction
with nearest neighbors. For other causes of death
we do not have even the beginning of an understanding.
\qbu Finally, a look at Fig. 6 and Table 4 shows that 
suicide is the cause of death for which the death ratios are
highest.
\qpar

\qA{Suicide ratios in France (1968-1993)}

%%-----------------------------------------------
%% MAPM#MARITAL9S + MAPM#MARITAL9W
\begin{figure}[htb]
\vskip -5mm
\centerline{\psfig{width=15cm,figure=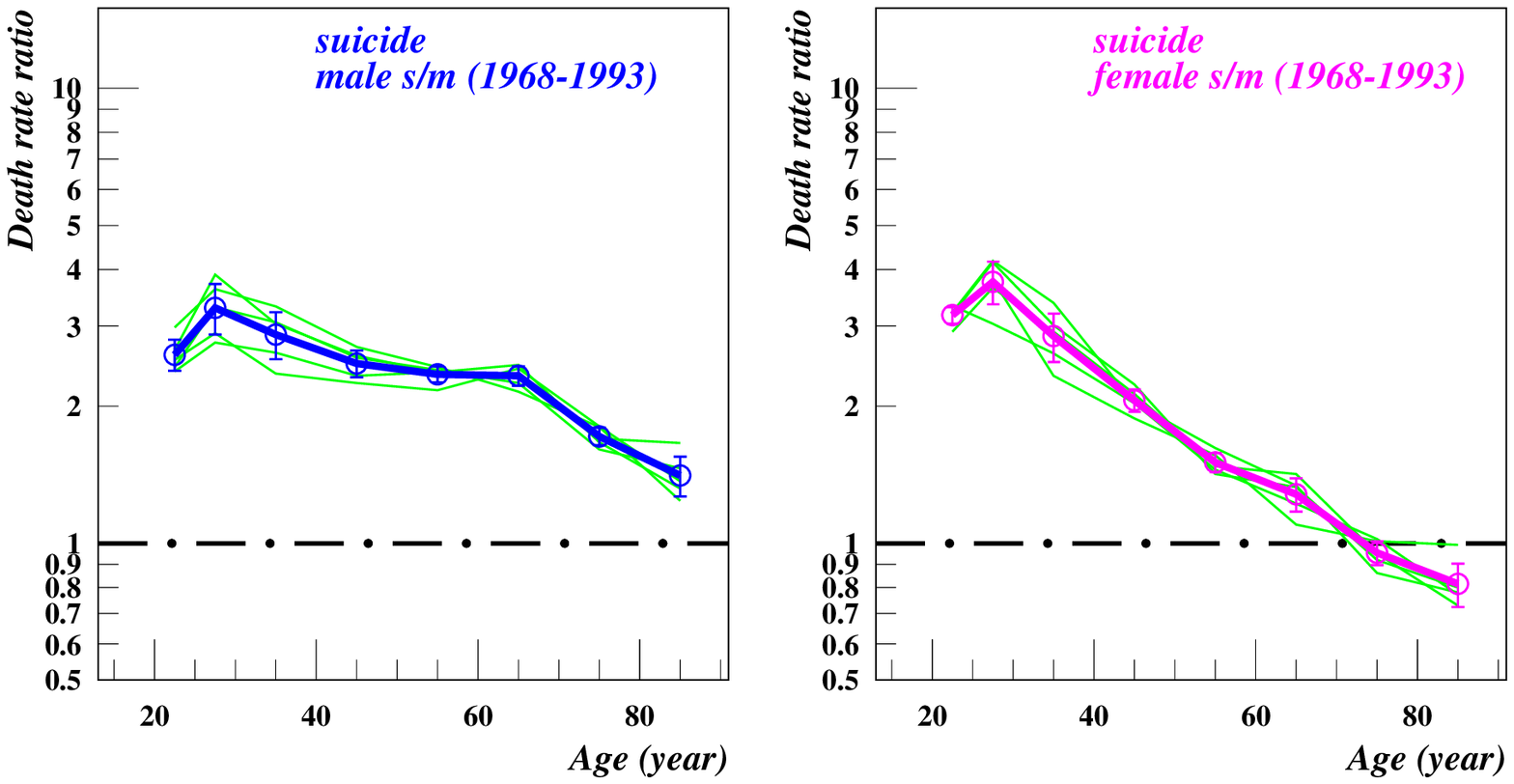}}
\vskip -5mm
\centerline{\psfig{width=15cm,figure=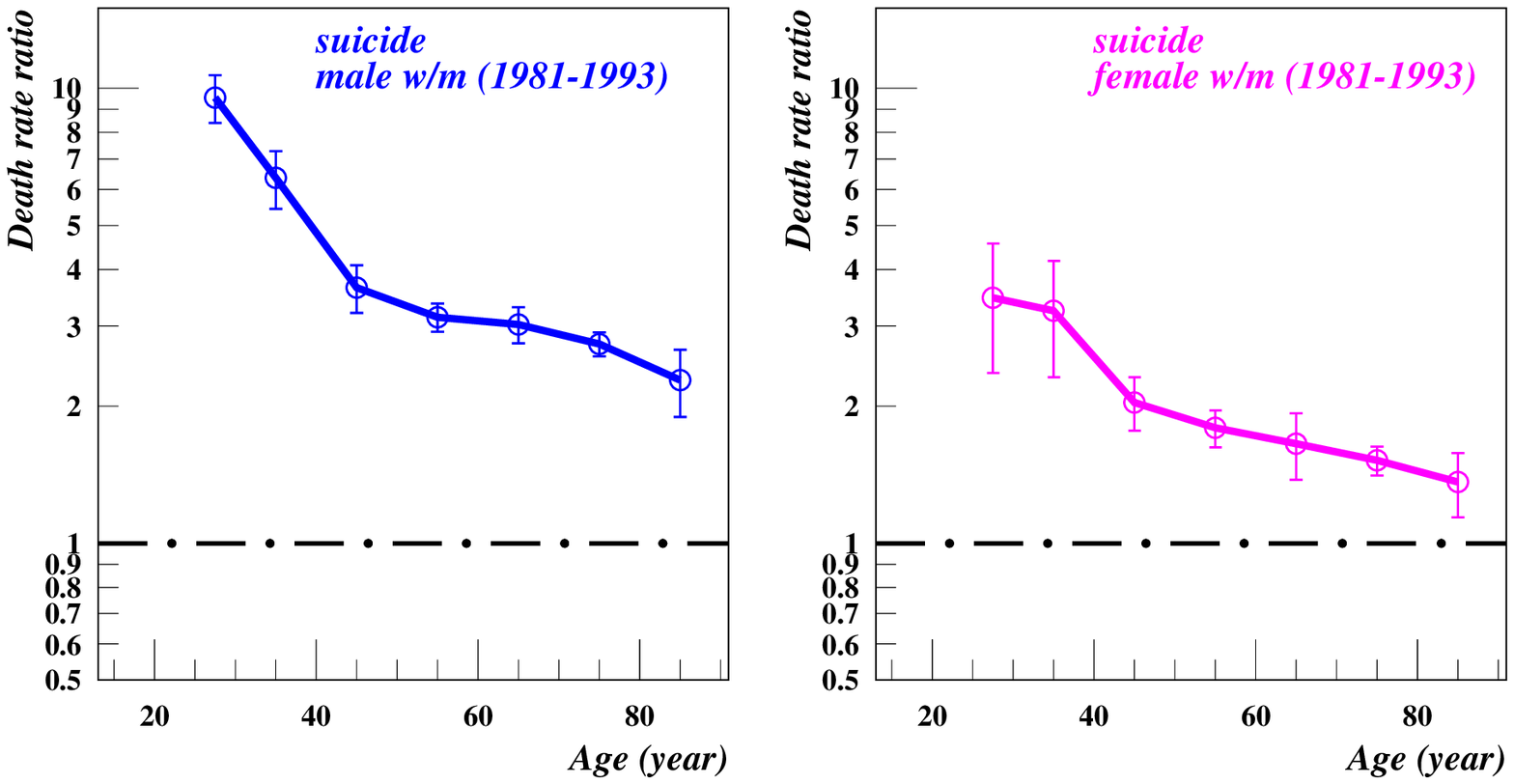}}
\qleg{Fig. 7a,b: Death rate ratio for suicide in
France.} 
{The age groups are: $ 20-24,25-29,30-39,40-49,
50-59,60-69,70-79,>80 $.}
{Source: Besnard (1997, p. 744, 752)}
\end{figure}
%-------------------------------------------------

In Fig. 6f we have fairly large statistical fluctuations.
As always in such a situation, we wish to reduce them.
There is only one way to do that: one must increase the
numbers of the events. This means either increasing the
size of the country or increasing the number of years.
Here we adopt the second approach.
We consider France,
which is smaller than the United States,
but for which data over a period of 26 years can be obtained.
The population of France is about
5 times smaller than the US population but 
with respect to the graphs displayed in Fig. 6f 
we will gain a factor: $  (26/2)/5=2.6 $.
In addition, shifting from the US to another country
will tell us something about the robustness of the
Farr-Bertillon effect.
\qpar

Regarding the accuracy of the present data set we must
also ask ourselves how the populations of the age-groups
by marital status have been estimated.
For inter-census years they were estimated through a computational
procedure whose results were published in Daguet (1995).
As already observed, the correctness of such a procedure
can hardly be checked. 

\qI{Conclusions and perspectives}

The Farr-Bertillon effect is not yet well recognized
as a major determinant of death rate. This is particularly
true for medical doctors and is attested by the fact that the 
characteristics of patients taking part in
medical trials include many parameters, yet their
marital status is usually not reported in spite
of the fact that it will substantially affect the
outcome of the trial. More details on this point can be found in
Roehner (2014).

\qA{Conclusions}

Our investigation of the Farr-Bertillon effect led to
the following conclusions.
\qbu In spite of the important demographic and 
sociological changes that took place over the past
century the Farr-Bertillon effect remained fairly unchanged.
\qbu The death rate ratios computed from different data sets
(based on censuses or surveys) are well consistent with one
another.

%%-----------------------------------------------
\begin{table}[htb]
\centerline{\bf Table 5\quad  Summary of death rate ratios $ w/m $ computed
in the present paper, USA, 1940-2010}

\vskip 3mm
\hrule
\vskip 0.7mm
\hrule
\vskip 1.2mm

\color{black} 
\small

$$ \matrix{
\hbox{Male} \hfill &\hbox{Age} & \hbox{\color{blue} Average} &
  \hbox{\color{blue} Average} &
\hfill \hbox{Coefficient} \cr
\hbox{or} \hfill &\hbox{}\hfill & \hbox{\color{blue} 1940-1960} &
\hbox{\color{blue} 1990-2010} &
\hfill \hbox{of}\cr
\hbox{Female} \hfill &\hbox{}\hfill& & \hbox{\color{blue} } &
\hfill \hbox{variation}\cr
\qtb
\hbox{} \hfill &\hbox{}\hfill && \hbox{\color{blue} } &
\hfill \hbox{for 1990-2010}\cr
\noalign{\hrule}
\qth
\hbox{Male} \hfill&\hfill && \hfill   & \hfill \cr
\hbox{} & 40  &  \color{blue} 3.2 &\color{blue} 4.5 & \hfill 2.6\% \cr
\hbox{} & 50  &  \color{blue} 2.3 &\color{blue}3.3 & \hfill 7.9\%\cr
\hbox{} & 60  &  \color{blue} 1.7 &\color{blue} 2.5& \hfill 12.0\%\cr
\hbox{Female} &\hfill && \hfill   & \hfill \cr
\hbox{} & 40  &   \color{blue} 2.2 &\color{blue}3.5 & \hfill 5.7\% \cr
\hbox{} & 50  &   \color{blue} 1.8 &\color{blue}2.7 & \hfill 8.6\% \cr
\qtb
\hbox{} & 60  &  \color{blue} 1.4 &\color{blue}2.5 & \hfill 11.1\% \cr
\noalign{\hrule}
}
$$
\vskip 2mm
Notes: The tables summarizes the death rate ratios $ w/m $ computed
in Fig. 3 for 1940-1960 and in Fig. 2, 4, 5 for 1990-2010.
The coefficient
of variation, that is to say $ \sigma/m $, gives an estimate
of the fluctuations due to various sources.
The age column gives the middle of the 10-year intervals
of the age groups.
The increase over past decades seen here
is also observed in Europe; the death rate ratios
$ (not-married)/m $ 
in several European  countries can be found in Vallin et al. (2001,
p. 318-321).
We do not yet know what are the factors
which bring about such an increase.
{\it }
\vskip 2mm
\hrule
\vskip 0.7mm
\hrule
\end{table}
%%-----------------------------------------------

The results summarized in Table 5 are restricted to the age interval
40-60 because Fig. 2 is limited to this interval.
\qbu Whereas for single and divorced persons the death rate
ratios are bell-shaped with a maximum around the age of 40,
for widowed persons it is a function which either decreases steadily
from youngest to oldest age groups or which has a maximum
at the second youngest age interval. It is this last shape which
is found in the observations of highest quality. 
However, we have seen that the shape of the curve for young 
age groups is mainly determined by the 
trafic accidents, a component which may be modified
by changes in trafic regulation rules.
\qbu When different causes of death are investigated it appears
that suicide leads to the highest average death rate ratio
while cancer leads to the lowest.
\qbu In recent decades death rate ratios have been increasing
in Europe as well as in the United States. We do not yet
know what is the reason of such an increase. 
May be it is a consequence of cohabitation without
marriage which has become so widespread in the US as in Europe
but it is not
clear how the two phenomena are related.
It should be observed that it is not only $ w/m $ which is 
increasing but also $ s/m $. As cohabitation makes 
the condition of ``singles'' (in mame) very similar to the 
condition of married persons one should observe a convergence od
$ d_s $ toward $ d_m $ that is to say a convergence of $ d_s/d_m $
toward 1.

\qA{Extraction of the dynamical response}

In fact, demographic statistics of the kind considered in the
present paper give little information about the dynamical
aspect of this phenomenon. We learned that ``on average''
for 10-year age-groups,
widowers have a death rate which is 3 times the death rate of
married persons. However, this observation does not tell us
anything about the transition from one state to another.
How long does it take? Does the death rate of recent
widowers increase steadily toward a steady state or is
there a shock effect during which the death rate 
ratio first overshoots its steady limit?
\qpar

One way to answer this question is to follow a sample of
married persons over several years. This was done in a
number of studies: Bojanovsky (1979, 1980),
Frisch et al. (2013), Helsing et al. (1981), 
Mellstr\"om et al. (1982), Parkes et al. (1969),
Thierry (1999, 2000), Young et al. (1963). 
Needless to say, in order to observe a substantial
number of deaths in a sample of married people 
followed by a sizable number of deaths 
of widowers one should
work on a sample of elderly persons. That is why
most of the previous studies concern persons over 50 or 60.
Basically, they found that the death rate of widowers
peaked in the first 3 or 6 months after widowhood
and then returned to the rate of married persons.
\qpar
How can one connect this observation to the
death rate ratios measured in the present paper?
Most of our observations concerned 10-year age groups?
Such age groups will contain a mixture of widowers who differ 
both in age ($ a $) and in the length of widowhood time
($ w $).
\qpar

Through the studies mentioned above, we know that the death rate
of widowers ($ d $) will be a function not only of $ a $ but
also of $ w $: $ d=d(a,w) $. Similarly, the remarriage rate
of widowers
($ d' $) will also be a function of $ a $ and $ w $: $ d'=d'(a,w) $.
Because age is recorded on death and marriage certificates,
one knows how, for fixed $ w $, $ d $ and $ d' $ depend upon
$ a $. Unfortunately, because $ w $ is {\it not} recorded on
death and marriage certificates, there is no direct
information about how $ d $ and $ d' $ depend upon $ w $ 
(for fixed $ a $). The functions $ d_a(w), d'_a(w) $ 
summarize the dynamic responses of a group of widowers of age $ a $.
\qpar
The objective of the second part of this study will be to extract
these functions from the data describing the Farr-Bertillon effect.
To say it in a different way, from the death rate ratio 
per age-group $ r_g=W_g/M_g $ we wish to extract $ r(w) $,
the {\it instantaneous} death rate ratio of a cohort who
entered widowhood at the same moment,
as a function of widowhood length.
\qpar

While being a well defined objective, it is not an easy one.
The solution that we will propose in Part 2 of this study may not
be the only one possible, nor may it be the simplest or most
satisfactory.
That is why we hope that the present paper will raise the
interest of other researchers. We 
welcome in advance any progress in this direction 
by other groups. That is also why the present paper has an
Appendix which provides
the primary data of those of our observations that can be
considered as the most accurate, namely the three stars observations of
Table 2. Those data should permit to test theoretical models.

\vskip 3mm

{\bf Acknowledgments}\quad The authors wish to express their
gratitude to
Drs. Jason Fields and Rose Kreider of the US Census Bureau,
and to Dr. Betzaida Tejada Vera of the Center for Diseases Control.
Their advice helped us to locate the most accurate data available.

\appendix

\qI{Appendix: Death rates by age, marital status and sex}

%%-----------------------------------------------
\begin{table}[htb]
%% CHIFFRES RECOPIES DE MAPM#MARITAL5
%
\centerline{\bf Table A1: Death rates by marital status and age,
USA, census years: 1940, 1950, 1960}

\vskip 3mm
\hrule
\vskip 0.7mm
\hrule
\vskip 1.2mm

\color{black} 
\small

$$ \matrix{
&<20&20&25&35&45&55&60&65&70&>75\cr
& &-&-&-&-&-&-&-&-&\cr
\qtb
& &24&34&44&54&59&64&69&74&\cr
\noalign{\hrule}
\qth
\hbox{\bf \color{blue} Men, death rate} \hfill& & & & & & & & & & \cr
\hbox{Single, 1940} \hfill
&4.5&2.9&4.7&9.2&17.4&28.7&38.3&52.3&75.1&132\cr
\hbox{Single, 1950}\hfill&3.3& 2.2& 3.6& 8.3& 17.2& 29.6& 40.8&
55.0& 79.5& 137\cr
\hbox{Single, 1960}\hfill 
&2.7&2.2&3.4&7.3&15.7&23.7&38.0&53.8&76.3&138\cr
\hbox{Married, 1940}\hfill 
&2.5 &2.2 &2.6 &4.8 &10.6 &19.1 &27.6 &39.4& 60.3 &113\cr
\hbox{Married, 1950}\hfill
&1.7 &1.4 &1.7 &3.6 &9.3 &17.7 &25.9 &36.6 &54.6 &100\cr
\hbox{Married, 1960}\hfill
&1.2 &1.2 &1.5 &3.0 &8.4 &16.0 &25.3 &37.2 &53.4 &100\cr
\hbox{Widowed, 1940}\hfill
&15.5 &11.8 &11.4 &14.1 &23.7 &34.8 &43.1 &57.4 &79.7 &162\cr
\hbox{Widowed, 1950}\hfill
&2.0 &6.1 &7.7 &12.3 &21.1 &30.2 &39.6 &49.7 &69.1&139\cr
\qtb
\hbox{Widowed, 1960}\hfill
&3.9 &5.4 &6.8 &10.6 &21.0 &32.0 &44.0 &57.9 &77.0 &156\cr
\noalign{\hrule}
}
$$
\vskip 2mm
Notes: The rates are given in deaths per 1,000 population of
specified group. Separate data for death numbers and populations
are not available. It can be observed that,
in accordance with Gompertz law (Gompertz 1825), the death rates
increase exponentially with age with a doubling time of about 10 years.\qL
{\it Source: Grove and Hetzel (1968, p. 334)}
\vskip 2mm
\hrule
\vskip 0.7mm
\hrule
\end{table}
%%-----------------------------------------------

%%-----------------------------------------------
\begin{table}[htb]
%% CHIFFRES RECOPIES DE MAPM#MARID2
%
\centerline{\bf Table A2: Death and population by marital status and age,
USA, census years: 1980, 1990, 2000}

\vskip 3mm
\hrule
\vskip 0.7mm
\hrule
\vskip 1.2mm

\color{black} 
\small

$$ \matrix{
\qtb
&\hfill 15-24&\hfill 25-34&\hfill 35-44&\hfill 45-54&\hfill
55-64&\hfill 65-74&\hfill >75\cr
\noalign{\hrule}
\qth
\hbox{\bf \color{blue} Men, deaths} \hfill& & & & & &\cr
\hbox{Single, 1980} \hfill
&\hfill30799 &\hfill13739 &\hfill6463 &\hfill10154 &\hfill16251 &\hfill20804
&\hfill25702\cr
\hbox{Single, 1990}\hfill
 &\hfill24921 &\hfill25448 &\hfill19324 &\hfill11609 &\hfill14492&\hfill
18973 &\hfill26611\cr
\hbox{Single, 2000}\hfill 
&\hfill21247 &\hfill16615 &\hfill21731 &\hfill20464 &\hfill15203
&\hfill18119 &\hfill28299\cr
\hbox{Married, 1980}\hfill 
&\hfill4894 &\hfill16144 &\hfill23238 &\hfill57550 &\hfill134153
&\hfill199962 &\hfill210401\cr
\hbox{Married, 1990}\hfill
&\hfill2386 &\hfill12868 &\hfill24234 &\hfill43802 &\hfill105347
&\hfill195495 &\hfill257631\cr
\hbox{Married, 2000}\hfill
&\hfill1493 &\hfill8218 &\hfill22160 &\hfill49298 &\hfill87351 &\hfill164106
&\hfil312544\cr
\hbox{Widowed, 1980}\hfill
&\hfill76 &\hfill223 &\hfill539 &\hfill2704 &\hfill12656 &\hfill38900
&\hfill130987\cr
\hbox{Widowed, 1990}\hfill
&\hfill29 &\hfill218 &\hfill619 &\hfill1852 &\hfill9329 &\hfill35630
&\hfill143605\cr
\hbox{Widowed, 2000}\hfill
&\hfill49 &\hfill148 &\hfill656 &\hfill2125 &\hfill7278 &\hfill30800
&\hfill182316\cr
\hbox{\bf \color{blue} Men, Population} \hfill& & & & & &\cr
\hbox{Single, 1980} \hfill
&\hfill17723 &\hfill4409 &\hfill978 &\hfill657 &\hfill551 &\hfill364
&\hfill198\cr
\hbox{Single, 1990}\hfill
&\hfill16516 &\hfill7779 &\hfill2493 &\hfill838 &\hfill555 &\hfill392
&\hfill225\cr
\hbox{Single, 2000}\hfill 
&\hfill17450 &\hfill7791 &\hfill4071 &\hfill1783 &\hfill548 &\hfill385
&\hfill247\cr
\hbox{Married, 1980}\hfill 
&\hfill3424 &\hfill12589 &\hfill10488 &\hfill9365 &\hfill7716 &\hfill5496
&\hfill2329\cr
\hbox{Married, 1990}\hfill
&\hfill2089 &\hfill12183 &\hfill13710 &\hfill9863 &\hfill8205 &\hfill6400
&\hfill3156\cr
\hbox{Married, 2000}\hfill
&\hfill2331 &\hfill10797 &\hfill15890 &\hfill13747 &\hfill9131 &\hfill6581
&\hfill4200\cr
\hbox{Widowed, 1980}\hfill
&\hfill8.05 &\hfill28.2 &\hfill53.3 &\hfill152 &\hfill366 &\hfill602 &\hfill
891\cr
\hbox{Widowed, 1990}\hfill
&\hfill10.1 &\hfill32.3 &\hfill71.8 &\hfill134 &\hfill346 &\hfill701
&\hfill1079\cr
\qtb
\hbox{Widowed, 2000}\hfill
&\hfill28.3 &\hfill120 &\hfill304 &\hfill665 &\hfill1776 &\hfill3587
&\hfill5637\cr
\noalign{\hrule}
}
$$
\vskip 2mm
Notes: The numbers of deaths are expressed in units while the
populations are in thousands. The data given in the table are the
primary data from which the death rate ratios $ s/m $ and
$ w/m $ were computed. The data for females can be drawn from the
same sources; they were omitted here in order to save space.\qL
{\it Sources: Same as for Fig. 4.}
\vskip 2mm
\hrule
\vskip 0.7mm
\hrule
\end{table}
%%-----------------------------------------------

This Appendix gives the death numbers and population data
for US census years from 1940 to 2000%
\qfoot{The regular publication of
death data by age, marital status and sex began in 1979.
However, because marital status was recorded on death certificates,
Grove and Hetzel (1968) from the US Census Bureau were able to 
compute and publish death rates for the
census years 1940, 1950 and 1960. That is why no death
numbers are available for these years. For 1970 there 
neither death rates nor death numbers.}%
. 
Although, most of these
data are available on Internet, they are not easy to
locate. For instance after 1970, census population data are buried
among dozens of volumes and thousands of pages
of census publications; this
makes their identification and extraction fairly time consuming.

\vskip 15mm

{\bf \large References} 

{\color{blue} The comments at the end of some of the entries
will be removed in the final version of the paper.
They focus on how those works contribute to
our present investigation.}
\qpar

\qparr
Accuracy of data on population characteristics as measured by
CPS-census match, 1960. Series ER 60, No 5. Bureau of the Census.
US Government Printing Office, Washington DC.  

\qparr
Bertillon (L.-A.) 1872: Article ``Mariage'' in the
Dictionnaire Encyclop\'edique des Sciences M\'edicales,
[Encyclopedic Dictionary of the Medical Sciences].
2nd series, Vol. 5, p.7-52.\qL
[Available on ``Gallica'', the website of digitized 
publications of the French national library, at http://www.bnf.fr]

\qparr
Bertillon (L.-A.) 1879: Article ``France'' in the
Dictionnaire Encyclop\'edique des Sciences M\'edicales,
[Encyclopedic Dictionary of the Medical Sciences].
4th series, Vol. 5, p.403-584.\qL
[Available on ``Gallica'', the website of digitized 
publications of the French national library at http://www.bnf.fr]

\qparr
Bertillon (J.) 1879: Les c\'elibataires, les veufs et les divorc\'es,
au point de vue du mariage. [Inclination to marriage 
respectively of bachelors, widowers and divorced persons.]
Revue Scientifique de la France et de l'Etranger 8,33,776-783.

\qparr
Besnard (P.) 1997: Mariage et suicide. la th\'eorie
durkheimienne de la r\'egulation conjugale \`a l'\'epreuve
d'un si\`ecle. [The impact of marriage on suicide revisited
one century after Durkheim's work].
Revue Fran\c{c}aise de Sociologie, 38,735-758.

\qparr
Bojanovsky (J.) 1979: Wann droht der Selbstmord bei Geschiedenen?
[At what point after separation are divorced people most
likely to commit suicide?]
Schweizer Archiv f\"ur Neurologie, Neurochirurgie und Psychiatrie
125,1,73-78.

\qparr
Bojanovsky (J.) 1980: Wann droht der Selbstmord bei Verwitweten?
[After becoming a widower when is the likelihood of committing
suicide largest?].
Schweizer Archiv f\"ur Neurologie, Neurochirurgie und Psychiatrie
127,1,99-103.

\qparr
Daguet (F.) 1995: Un si\`ecle de d\'emographie fran\c{c}aise.
Structure et \'evolution de la population de 1901 \`a 1993.
[Structure and evolution of the French population from
1901 to 1993], Institut National de la Statistique et des
Etudes Economiques (INSEE), No 434-435, Paris.

\qparr
Deaths: final data. National Vital Statistics Reports, 
Years 1996-2010.
National Center for Health Statistics,
Hyattsville, Maryland.\qL
[available on the website of the NCHS at: \qL
http://www.cdc.gov/nchs/products/nvsr.htm]

\qparr
Durkheim (E.) 1897: Le suicide. Etude de sociologie. F. Alcan, Paris.
A recent English translation is: ``On Suicide'' (2006),
Penguin Books, London.\qL
[Durkheim showed not only that 
suicide rates among non-married, widowed or divorced
persons were higher than among married persons but also
that they were lower among married persons with several
children than for married persons without children.]

\qparr
Farr (W.) 1859, 1975: Influence of marriage on the mortality of the
French people (12 p.). Transactions of the National Association
for the Promotion of Social Science 1858-1859, 504-520.\qL
The paper was republished in 1975 in ``Vital statistics, a memorial
volume of selections from reports and writings of William Farr''.
Scarecrow Press, Methuen (New York).
[The style of the report is somewhat outdated and sometimes confusing.
An illustration is provided
by the following excerpt taken from the first paragraph.
``The action of the various parts of the body in industrial occupations
  produces specific effects. Every science modifies its
  cultivators. The play of the passions transfigures the human
  frame. How do they influence its existence?'']

\qparr
Flounders (S.) 2009: Why U.S. occupation cannot transform Afghanistan
or Iraq. [available on the website of the ``International
Action Center'']

\qparr
Frisch (M.), Simonsen (J.) 2013: Marriage, cohabitation and mortality
in Denmark: national cohort study of 6.5 million persons followed
for up to 3 decades. International Journal of Epidemiology 1,13.

\qparr
Gompertz (B.) 1825: On the nature of the function expressive of
the law of human mortality, and on a new mode of determining the value
of life contingencies. Philosophical Transactions of the Royal
Society 115,513–585. 

\qparr
Gove (W.R.) 1972: Sex, marital status, and suicide. Journal of
Health and Social Behavior 13,204-213.

\qparr
Grove (R.D.), Hetzel (A.M.) 1968:  Vital statistics rates 
in the United States, 1940-1960. United States Printing Office,
Washington DC. 

\qparr
Helsing (K.J.), Szklo (M.), Comstock (G.W.) 1981: Factors 
associated with mortality after widowhood. 
American Journal of Public Health 71,802-809.\qL
[The paper is based on the comparison of a sample of 4,032 persons
who became widowed between 1963 and 1974 and a sample
of same size of married persons. For males, the ratio of death
rates of widowed persons to that of married persons is 1.34.
The paper is freely available on the Internet.]

\qparr
Koposova (A.J.), Breault (K.D.), Singh (G.K.) 1995:
White man suicide in the United States. A multivariate
individual-level analysis. Social Forces, 74,1,315-323.

\qparr
Linder (F.E.), Grove (R.D.) 1947: Vital statistics rates 
in the United States, 1900-1940. United States Printing Office,
Washington DC.

\qparr
March (L.) 1912: Some researches concerning the factors of 
mortality. Journal of the Royal Statistical Society 75,505-538.\qL
[The paper includes a comparative analysis of mortality rates
in France, Prussia and Sweden.]

\qparr
Mellstr\"om (D.), Nilsson (A.), Od\'en (A.), Rundgren (A.), 
Svanborg (A.) 1982: Mortality among the widowed in Sweden.
Scandinavian Journal of Social Medicine 10,33-41.\qL
[The study followed
360,000 individuals aged between 50 and 90
who were widowed in Sweden at some point between 1968 and 1978. 
It showed a peak in the mortality risk during the first 3 months.
For widowers the amplitude (with respect to married persons) was 1.48
while for widows there was a peak of smaller amplitude, namely 1.22.]

\qparr
Mortality statistics: review of the Registrar General on deaths
in England and Wales. Series DHI, Number 16. Her Majesty's Stationary
Office, London. 

\qparr
National Center for Health Statistics (NCHS) 1970: Mortality
from selected causes by marital status. Vital and Health
Statistics, Series 20, number 8.

\qparr
Parkes (C.M.), Benjamin (B.), Fitzerald (R.G.) 1969:
Broken heart. A statistical study of increased mortality among
widowers. British Medical Journal 1,740-743.
[This is the continuation of the Young et al. (1963) study.
The same sample was observed over 4 more years i.e. a total
of 9 years after the death of the spouse. Over these 4 years
the death ratio widowed/married was comprised between 0.90 and
0.95.]

\qparr
Raff (M.C.) 1998: Cell suicide for beginners. Nature 396, 12 November,
119-122.

\qparr
Registrar General 1971: Statistical Review of England and Wales.
Part III. Office of Population Censuses and Surveys. London.

\qparr 
Roehner (B.M.) 2007: Driving forces in physical, biological
and socio-economic phenomena. Cambridge University Press,
Cambridge.

\qparr
Roehner (B.M.) 2014: Incidence of the Bertillon and Gompertz effects
on the outcome of clinical trials. Physica A, 414, p. 300-307.

\qparr
Statistisches Jahrbuch f\"ur die Bundesrepublik Deutschland 1978:
Stuttgart.

\qparr
Stroebe (W.), Stroebe (M.S.) 1987: Bereavement and health.
The psychological and physical consequences of partner loss.
Cambridge University Press, Cambridge.\qL
[The book is mostly concerned with psychological and
other qualitative aspects. Although short (p. 151-167), the
review of the quantitative evidence about the mortality of widowed
persons is quite useful. It can be noted that the book has
a broad reference section which contains about 600 entries.]

\qparr
Thierry (X.) 1999: Risques de mortalit\'e et de surmortalit\'e au
cours des 10 premi\`eres ann\'ees de veuvage.
[Excess mortality during the first 10 years of widowhood.] 
Population 54,2,177-204. 

\qparr
Thierry (X.) 2000: Risques de mortalit\'e et causes m\'edicales
des d\'ec\`es aux divers moments du veuvage. 
G\'erontologie et Soci\'et\'e 95,27-43.

\qparr
Vallin (J.), Mesl\'e (F.), Valkonen (T.) 2001: 
Tendances en mati\`ere de mortalit\'e et mortalit\'e diff\'erentielle.
Editions du Conseil de l'Europe, Strasbourg.\qL
An English version was published under the title
``Trends in mortality and differential mortality''.

\qparr
Wang (L.), Xu (Y.), Di (Z.), Roehner (B.M.) 2013:
How does group interaction and its severance affect life expectancy? 
arXiv preprint 1304.2935 (9 April 2013) 

\qparr
Young (M.), Benjamin (B.), Wallis (C.) 1963: Mortality of
widowers. Lancet 2,254-256.\qL
[This is a longitudinal study.
The authors followed over a period of 5 years
a sample of 4,486 widowers more than 55 years old and
whose wives had died in January 1957 . They found a 
death rate ratio widowed/married of 1.39 in the first six months after
the death of the spouse and 1.04 for the remaining time. \qL
The problem
is that these ratios are too low to account for the ratio
of 1.6 given by UK vital statistics data for 
age groups. On average individuals
spend 5 years in a 10-year age group. If their death rate is 
multiplied by $ k_1 $ over an interval of $ m $ months, 
and by $ k_2 $ for the remaining months, then
for the 5 years during which they remain in the age group
the average rate will be multiplied by:
 $ k=m\times k_1 + (60-m)\times k_2)/60 $;
for $ m=6, k_1=1.39, k_2=1.04 $ one gets: $ k=1.08 $.]

\end{document}